\documentclass[acmtog]{acmart}
\usepackage{subfiles}
\usepackage{enumitem}
\usepackage{relsize}
\usepackage{mathtools}

\setcopyright{none}
\acmDOI{}
\renewcommand\footnotetextcopyrightpermission[1]{}

\newcommand*\unit[1]{\, \mathrm{#1}}

\newcommand{\dquote}[1] {``#1''}
\newcommand{\pystorms}{\texttt{pystorms} }
\newcommand{\pystormsNOSPACE}{\texttt{pystorms}}

\newcommand{\scenario}[1]{\texttt{#1}}
\newcommand{\code}[1]{\texttt{#1}}

\newcommand{\demotheta}{\href{https://colab.research.google.com/github/kLabUM/pystorms/blob/master/tutorials/Scenario_Theta.ipynb?hl=en}{pystorms.org/\#ipynb/GoogleColab/theta}}
\newcommand{\demoalpha}{\href{https://colab.research.google.com/github/kLabUM/pystorms/blob/master/tutorials/Scenario_Alpha.ipynb?hl=en}{pystorms.org/\#ipynb/GoogleColab/alpha}}
\newcommand{\demobeta}{\href{https://colab.research.google.com/github/kLabUM/pystorms/blob/master/tutorials/Scenario_Beta.ipynb?hl=en}{pystorms.org/\#ipynb/GoogleColab/beta}}
\newcommand{\demogamma}{\href{https://colab.research.google.com/github/kLabUM/pystorms/blob/master/tutorials/Scenario_Gamma.ipynb?hl=en}{pystorms.org/\#ipynb/GoogleColab/gamma}}
\newcommand{\demodelta}{\href{https://colab.research.google.com/github/kLabUM/pystorms/blob/master/tutorials/Scenario_Delta.ipynb?hl=en}{pystorms.org/\#ipynb/GoogleColab/delta}}
\newcommand{\demoepsilon}{\href{https://colab.research.google.com/github/kLabUM/pystorms/blob/master/tutorials/Scenario_Epsilon.ipynb?hl=en}{pystorms.org/\#ipynb/GoogleColab/epsilon}}
\newcommand{\demozeta}{\href{https://colab.research.google.com/github/kLabUM/pystorms/blob/master/tutorials/Scenario_Zeta.ipynb?hl=en}{pystorms.org/\#ipynb/GoogleColab/zeta}}

\begin{abstract}  
Recent accessibility of affordable sensing technologies, microcontrollers, and wireless communication technology has made it possible for stormwater systems to be retrofitted with an assortment of sensors and actuators. 
These \textit{smart stormwater systems} have enabled the real-time sensing of their surrounding environmental dynamics, and subsequently, provide the basis for autonomous and adaptive operational control strategies. 
Additionally, these systems allow for inexpensive and minimally-invasive stormwater control interventions (e.g. hydraulic valve operated by cellularly-connected actuator) in lieu of new construction.
However promising this area of smart stormwater control, there still remain barriers --- for experts and novices alike --- to access a set of shared tools and methods for investigating, developing, and contributing to it.
In an effort to make smart stormwater control research more methodical, efficient, and accessible, we present \pystormsNOSPACE, a Python-based simulation sandbox that facilitates the quantitative evaluation and comparison of control strategies.
The \pystorms simulation sandbox comes with (i) a collection of real world-inspired stormwater control scenarios on which any number of control strategies can be applied and tested via (ii) an accompanying Python programming interface coupled with a stormwater simulator. 
For the first time, \pystorms enables rigorous and efficient evaluation of smart stormwater control methodologies across a diverse set of watersheds with only a few lines of code.
We present the details of \pystorms here and demonstrate its usage by applying and evaluating two stormwater control strategies.
\end{abstract}
\begin{document}
%
%
\title{\pystormsNOSPACE:\ A simulation sandbox for the development and evaluation of stormwater control algorithms}
\author{Sara P. Rimer}
\authornote{These authors contributed equally to the paper.}
\authornote{Corresponding authors}
\orcid{0000-0002-3035-7698}
\affiliation{\institution{Argonne National Laboratory}
    \department{Decision and Infrastructure Sciences Division}
	\streetaddress{9700 Cass Avenue}
	\city{Lemont}
	\state{IL}
	\postcode{60439}
	\country{USA}
}
\email{srimer@anl.gov}	 
\author{Abhiram Mullapudi}
\authornotemark[1]
\authornote{Present affiliation: Xylem, Inc., South Bend, IN USA}
\orcid{0000-0001-8141-3621}
\affiliation{\institution{University of Michigan}
    \department{Civil and Environmental Engineering Department}
	\streetaddress{2350 Hayward Avenue}
	\city{Ann Arbor}
	\state{MI}
	\postcode{48105}
	\country{USA}
}
\email{abhiram.mullapudi@xylem.com}
\author{Sara C. Troutman}
\authornotemark[1]
\authornotemark[3]
\orcid{0000-0002-6809-7959}
\affiliation{\institution{University of Michigan}
    \department{Civil and Environmental Engineering Department}
	\streetaddress{2350 Hayward Avenue}
	\city{Ann Arbor}
	\state{MI}
	\postcode{48105}
	\country{USA}
}
\email{sara.troutman@xylem.com}
\author{Gregory Ewing}
\orcid{0000-0002-0106-7712}
\affiliation{\institution{University of Iowa}
    \department{Department of Civil and Environmental Engineering}
	\streetaddress{300 South Riverside Drive}
	\city{Iowa City}
	\state{IA}
	\postcode{52245}
	\country{USA}
}
\email{gregory-ewing@uiowa.edu}
\author{Benjamin D. Bowes}
\orcid{0000-0001-8349-4787}
\affiliation{\institution{University of Virginia}
    \department{Department of Engineering Systems and Environment}
	\streetaddress{151 Engineer's Way}
	\city{Charlottesville}
	\state{VA}
	\postcode{22904}
	\country{USA}
}
\email{bdb3m@virginia.edu}
\author{Aaron A. Akin}
\authornote{Present affiliation: Davis \& Floyd, Charleston, SC USA}
\orcid{0000-0001-7033-9307}
\affiliation{\institution{University of Tennessee}
    \department{Department of Civil and Environmental Engineering}
	\streetaddress{851 Neyland Drive}
	\city{Knoxville}
	\state{TN}
	\postcode{37996}
	\country{USA}
}
\email{aakin@davisfloyd.com}
\author{Jeffrey Sadler}
\authornote{Present affiliation: United States Geological Survey, Middleton, WI USA}
\orcid{0000-0001-8776-4844}
\affiliation{\institution{University of Virginia}
    \department{Department of Engineering Systems and Environment}
	\streetaddress{151 Engineer's Way}
	\city{Charlottesville}
	\state{VA}
	\postcode{22904}
	\country{USA}
}
\email{jms3fb@virginia.edu}
\author{Ruben Kertesz}
\orcid{0000-0001-7936-8235}
\affiliation{\institution{Xylem, Inc.}
	\streetaddress{121 South Niles Avenue, 32}
	\city{South Bend}
	\state{IN}
	\postcode{46617}
	\country{USA}
}
\email{ruben.kertesz@xylem.com}
\author{Bryant McDonnell}
\orcid{0000-0002-6250-2220}
\affiliation{\institution{Xylem, Inc.}
	\streetaddress{121 South Niles Avenue, 32}
	\city{South Bend}
	\state{IN}
	\postcode{46617}
	\country{USA}
}
\email{bryant.mcdonnell@xylem.com}
\author{Luis Montestruque}
\orcid{0000-0001-5494-585X}
\affiliation{\institution{Xylem, Inc.}
	\streetaddress{121 South Niles Avenue, 32}
	\city{South Bend}
	\state{IN}
	\postcode{46617}
	\country{USA}
}
\email{luis.montestruque@xylem.com}
\author{Jon Hathaway}
\orcid{0000-0003-1666-8550}
\affiliation{\institution{University of Tennessee}
    \department{Department of Civil and Environmental Engineering}
	\streetaddress{851 Neyland Drive}
	\city{Knoxville}
	\state{TN}
	\postcode{37996}
	\country{USA}
}
\email{hathaway@utk.edu}
\author{Jonathan L. Goodall}
\orcid{0000-0002-1112-4522}
\affiliation{\institution{University of Virginia}
    \department{Department of Engineering Systems and Environment}
	\streetaddress{151 Engineer's Way}
	\city{Charlottesville}
	\state{VA}
	\postcode{22904}
	\country{USA}
}
\email{goodall@virginia.edu}
\author{Branko Kerkez}
\authornotemark[2]
\orcid{0000-0002-8041-5366}
\affiliation{\institution{University of Michigan}
    \department{Civil and Environmental Engineering Department}
	\streetaddress{2350 Hayward Avenue}
	\city{Ann Arbor}
	\state{MI}
	\postcode{48105}
	\country{USA}
}
\email{bkerkez@umich.edu}
%
%
%
%
%
%
\maketitle
%
%
%
%
%
%
%
%
%
%
%
%
%
%
\section{Introduction}
\label{sec:introduction}
%
%
%
The advent of smart cities is poised to transform the management of our built environment \citep{Harrison2011, Batty2012, Chourabi2012, Kitchin2014, Bibri2017, Eggimann2017}. Specific to stormwater, a new generation of smart and connected stormwater systems aims to reduce flooding and improve water quality management by autonomously sensing watershed parameters and controlling hydraulic components across complete watersheds. These smart systems can provide an alternative to costly concrete-and-steel construction by squeezing even more performance out of existing stormwater infrastructure. Early ideas of controlling distributed stormwater systems in real-time date back to the 1970s \citep{Trotta1977}. The concept has, however, only recently gained widespread attention---in large part due to the affordability of internet-connected sensors, the increased capacity of data services, the broader emergence of other autonomous systems such as self-driving cars and robots, and the increasing prevalence of climate change stressors such as changing rainfall patterns and sea level rise. Relative to other fields of autonomy, however, smart water systems are still in the early stages of adoption. Thus, developing and implementing smart water systems presents an exciting opportunity for researchers and practitioners to propose new visions, standards, and technologies.

 \
 
The intelligence of smart stormwater systems broadly refers to the acquisition (i.e. \dquote{sensing}) and processing of data into decisions and actions (i.e. \dquote{control strategies}) that are then used to guide the operation of gates, valves, pumps, and other actuators within a watershed or drainage network. Ultimately, the logic embedded via these control rules determines how water is moved around the collection system to meet specific performance objectives, such as the reduction of flooding or improvement of water quality. Developing this control logic poses a major research frontier \citep{Kerkez2016} and will require the engagement of groups and individuals from a wide range of experience and expertise. 

\

Yet, entering this research field is presently precluded by a number of practical barriers. For new groups to make early contributions, they must have access to simulation testbeds, real-world inspired case studies, and appropriate control objectives. Due to privacy and safety concerns, access to network models and management details is difficult to come by without personal connections to those who manage local watersheds or municipal water systems. Thus, it can be difficult for new groups to obtain the necessary details of how real-world stormwater systems actually operate---leaving them to evaluate their control algorithms solely on idealized ``toy problems." When access to such details is available, developing computational simulations that are true to real-world systems and objectives requires significant effort and expertise. Still, even when all of these other barriers are addressed and promising control algorithms have been proposed, they have usually all been evaluated on highly specific case studies and simulators, making it difficult to evaluate the extent of their success when applied to additional networks. In an effort to address these limitations, the contribution of this paper is \pystormsNOSPACE, an open-source Python package comprised of:
\begin{enumerate}[label=(\roman*)]
\item A collection of real world-inspired smart stormwater control scenarios that facilitate the quantitative evaluation and comparison of control strategies.
\item A programming interface and a stormwater simulator to provide a stand alone package for developing stormwater control strategies.
\end{enumerate}
Our aspiration is for \pystorms to emerge as a community-driven resource that fosters accessibility and collaboration amongst researchers and practitioners, both novices and experts alike. 
%
%
%
%
%
%
%
%
%
%
%
%
%
%
%
\section{Background}
\label{sec:background}
%
%
%
\begin{figure*}
    \centering
    \includegraphics[width=\linewidth]{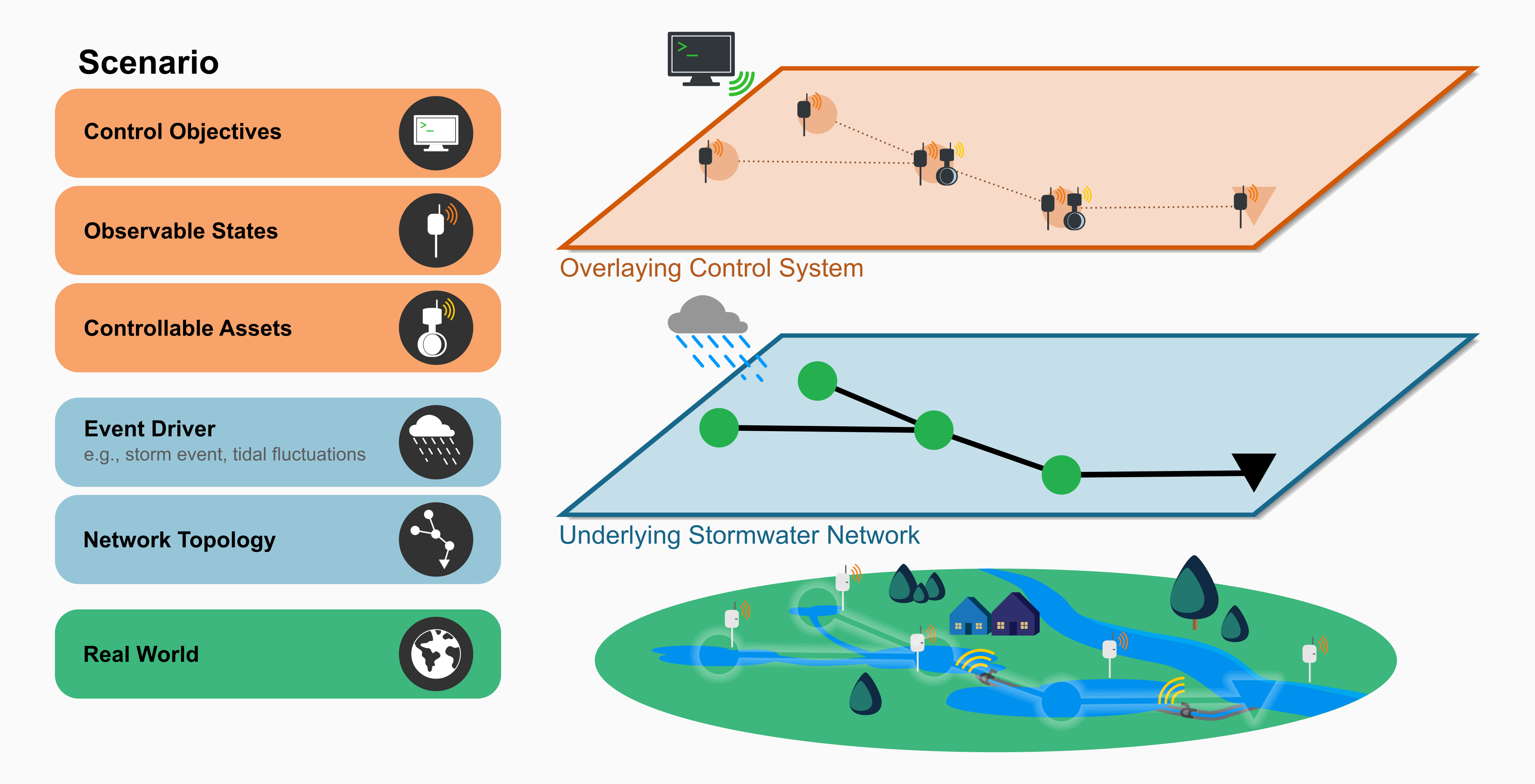}
    \caption{\pystorms abstracts the control of stormwater systems as scenarios, each characterized by an overlaying control system and an underlying stormwater network.
    The overlaying control system (in orange) encapsulates what can be considered the stormwater control system's virtual elements --- that is, the components of the stormwater control system that are changeable and/or readable by an implemented control algorithm --- and include its Control Objectives, Observable States, and Controllable Assets.
    The underlying stormwater network (in blue) represents the computational implementation of the hydraulic and hydrological elements of the stormwater control system, namely its Network Topology and Event Driver.
    Table \ref{tab:terminology} presents a detailed description of these elements.}
    \label{fig:scenarioComponents}
\end{figure*}
\subsection{Control of Stormwater Systems}
\label{subsec:controlofstormwatersystems}
A stormwater control problem can be defined as finding a strategy to manipulate the flow of stormwater to achieve a desired water quantity or quality objective.
Traditionally, stormwater control has relied on \emph{passive} solutions, in which control is achieved via large-scale and expensive construction. 
Instead, the emergence of microcontrollers, wireless communication technologies, and low-cost sensors has allowed for \emph{active} solutions, in which existing infrastructure can be retrofitted with low-cost wireless valves, pumps, and other actuators, installed at strategic locations throughout a stormwater network, and utilized adaptively, in \emph{real-time}. Consequently, stormwater infrastructure can now be instantly redesigned to respond to its dynamic environment.

\

While real-time stormwater control engineering solutions were documented at least a decade earlier \citep{Trotta1977}, research-oriented discussion of these implementations did not occur until 1989 \citep{Schilling1989}. Furthermore, while limited implementation of smart stormwater control began at the end of the 20th-century, the 21st-century has seen far more extensive and systematic successes, as described in the foundational reviews of \citet{Schutze2004} and \citet{Vanrolleghem2005}. Some notable smart stormwater control implementations include \citet{Ocampo-Martinez2010, Gaborit2013, Vezzaro2014, Garcia2015, Gaborit2016, Mullapudi2017, Montestruque2018, Sadler2019, Persaud2019, Bowes2021}. These references contain instances of smart stormwater control from single control assets to watershed-scale implementations.
For more comprehensive reviews of stormwater control implementations, we direct the reader to some recently published survey articles on the topic \citep{Yuan2019, Lund2018, Shishegar2018, VanDaal2017, Garcia2015}. 
\subsubsection{Simulating Stormwater Systems}
\label{subsec:simulatingstormwatersystems}
Due to safety concerns and the variability of storm events, it is often infeasible to test various control strategies on actual stormwater networks. A more practical approach is to first estimate the outcomes of different control decisions \emph{in-silico}, before deciding to port them to real systems. These simulations can be carried out using a computational stormwater model. The components of such a model usually include a (i) runoff module and (ii) a routing module, and are driven by (iii) precipitation events (e.g. rain, snow). The runoff module converts precipitation into overland runoff; the overland runoff then undergoes hydrological processes (e.g. infiltration, evaporation) and is hydraulically transported to the stormwater collection system.

\

Over the years, several different software applications have been developed for modeling and simulating stormwater networks. The different software applications all function in a similar manner: they compute the dynamics of stormwater as it moves through its local watershed. The US-EPA's Stormwater Management Model (SWMM) \citep{Rossman2015a}, MIKE URBAN+ from the MIKE Powered by DHI software suite\footnote{\href{https://www.mikepoweredbydhi.com/products/mike-urban-plus}{mikepoweredbydhi.com}}, and the Model for Urban Stormwater Improvement Conceptualisation (MUSIC) by eWater\footnote{\href{https://ewater.org.au/products/music/}{ewater.org.au}} are examples of popular stormwater software applications. The theory and computational details of these models are summarized in \citet{Rossman2015b}, \citet{Rossman2017}, and \citet{Rossman2016}. While these models provide powerful simulation adroitness, they confine real-time control to limited rule-based approaches, making them difficult to use in the study of smart stormwater systems. %
\subsubsection{Implementing Adaptive Control}
\label{subsec:implementingcontrol}
Here, we aim to maintain the idea of control in its broadest but most straightforward meaning: after receiving some sort of \emph{cue} (from a sensor or state estimate), an \emph{action} is taken (for instance opening or closing a valve) with the aim of achieving a desired outcome (such as reducing flooding or improving water quality). When we implement \emph{control}, we are deciding on the course of action for optimizing our system to meet some specified objective. A stormwater control strategy seeks to formalize this process of deciding a set of actions. We define the computational process of implementing this strategy as the \emph{stormwater control algorithm}.

\

Suppose a stormwater system with only one valve was installed, and that valve can either be completely opened or closed every hour. Deciding on a pattern for the complete opening and closing of the valve, over the period of a few hours to even a few days, presents a multitude of permutations and is a non-trivial undertaking. By expanding the task to allow for the valve to opened at any number of percent increments between 0--100\%, the combination of actions that can be implemented becomes even more expansive, making simple rule-based or \dquote{if-else} algorithms either ineffective or bafflingly large. This complexity is particularly true for more sizable drainage systems, where tens to hundreds of valves need to be operated. 

\

Hence, finding the \dquote{best} control actions is difficult and---by nature---subjective. Furthermore, most algorithms are often only evaluated on a single case study, and only in comparison to an uncontrolled solution. While valuable, this has the effect of making most algorithms appear very effective, since the only baseline that exists otherwise is the unoptimized, \dquote{as-built} static system. 
\subsection{The Need for a Simulation Sandbox}
\label{subsec:overviewofthesimulationsandbox}
For new researchers and practitioners in the field of smart stormwater control, the barrier to entry is significant. Newcomers must spend considerable time searching for real-world case studies, synthesizing relevant control objectives, developing algorithms, and building the capacity to carry out simulation in order to test new ideas. To grow this research field and support its broader scientific context, there is a need to better enable the cross-comparison of smart stormwater control strategies, their algorithms, and the case study instances for when they are used. While there have been some prior efforts to benchmark specific stormwater networks in order to evaluate  adaptive control strategies \citep{Schutze2017, Borsanyi2008}, there is still a shortage of stormwater control case studies with diverse control objectives.

\

Other research domains provide compelling examples of community-driven simulation tool chains that have been developed for similar cross-comparison needs. For example, the ARPA-E GRID DATA program has enabled an active research community in the energy sector by providing open source case studies and benchmarking tools of power system networks \citep{Griddata2016}. In a similar vein, the water distribution community has created their own active cross-comparison tools \citep{Walski1987, Ostfeld2008, Marchi2014}. \pystorms is our effort to develop a similar set of research tools for the stormwater control community. Additionally, we have also been inspired by the streamlined control algorithm testing of the OpenAI Gym toolkit \citep{Brockman2016}. As such, we contend that there is a need for a similar, more \dquote{out-of-the-box} software toolset with an unambiguous programming interface that allows stormwater control researchers to get started more quickly. To that end, we have formulated \pystorms as a \emph{simulation sandbox}, in which we systematize a collection of stormwater control testbed examples, and foster the experimentation and testing of new control strategies. 
\begin{table*}[ht]
\small
\caption{A \pystorms scenario is comprised of five distinct components: its Network Topology, an Event Driver, a set of Controllable Assets, a set of Observable States, and the Control Objectives.}
\label{tab:terminology}
\begin{tabular}{l p{5.5in}}
\toprule
\textbf{Network Topology} & A \emph{network} is the physical system of conduits (e.g.\ pipes, culverts), storage elements (e.g.\ retention and detention basins), and any other subcatchment infrastructure (e.g.\ green infrastructure, wetlands) that collect, convey, and/or treat stormwater runoff. \\\midrule
\textbf{Event Driver} & The \emph{event driver} consists of any inputs or \dquote{disturbances} to the network that govern the generation and flow of runoff.
Most often, an event driver is the precipitation generating runoff in the watershed, but can also include wastewater flows, tidal fluctuations, or other phenomena. \\\midrule
\textbf{Controllable Assets} & The \emph{controllable assets} are the subset of the network topology that are equipped with valves, pumps, or any other flow control infrastructure that can be actuated to manipulate stormwater flow. \\\midrule
\textbf{Observable States} & The \emph{observable states} are the collection of states in the network that can be accessed by the users during a simulation. In the real world, these states are measured by a set of sensors installed at the corresponding network locations. \\\midrule
\textbf{Control Objectives} & The overall goal or set of goals (e.g.\ preventing flooding,  reducing erosion, minimizing overflows, improving water quality) of manipulating the behavior of a stormwater network using controllable assets during a simulation. A control strategy's ability to achieve a particular objective is quantified via a corresponding performance metric.
\\
\bottomrule
\end{tabular}
\end{table*}
%
%
%
%
%
%
%
\section{\pystorms}\label{sec:pystorms}
%
%
%
To achieve the objectives of curating an open repository of smart stormwater control testbed examples, and reducing the learning curve for newcomers testing new control strategies, \pystorms is underpinned by two distinct features.
First, it provides a collection of diverse stormwater control scenarios, which are drawn from real-world urban watersheds to encompass diverse features appertaining to stormwater systems (Section \ref{subsec:scenarios}). Second, these scenarios are coupled with a streamlined Python programming interface (Section \ref{subsec:api}) that explicates the computational backend of a corresponding stormwater control simulator (Section \ref{subsec:architecture}). Together, these features provide researchers with a standalone software package that focuses its usage on the development and testing of stormwater control algorithms. 
\subsection{Scenarios}
\label{subsec:scenarios}
\begin{table*}[ht]
\small
\caption{\pystorms includes a curated collection of real world-inspired stormwater scenarios.}\label{tab:scenarios}
\begin{tabular}{p{0.5in}p{1.75in}p{1.75in}}
\toprule
\textbf{Scenario} & \textbf{Network Topology} & \textbf{Control Objectives} \\\midrule
\scenario{theta} & $2 \unit{km^2}$ idealized separated stormwater network & Maintain the flows at the outlet below a threshold and avoid flooding (2 storage basin outlets) \\\midrule
\scenario{alpha} & $0.12 \unit{km^2}$ residential combined sewer network & Minimize total combined sewer overflow volume (5 weirs at interceptor connections) \\\midrule
\scenario{beta} & $1.3 \unit{km^2}$ separated stormwater network with a tidally-influenced receiving river & Minimize flooding (1 storage basin outlet, 1 pump, 1 inline storage dam) \\\midrule
\scenario{gamma} & $4 \unit{km^2}$ highly urban separated stormwater network & Maintain channel flows below threshold and avoid flooding (11 detention pond outlets) \\\midrule
\scenario{delta} & $2.5 \unit{km^2}$ combined sewer network in which the stormwater ponds also serve as waterfront & Maintain water levels within upper and lower thresholds for water quality and aesthetic objectives (4 storage basin outlets; 1 infiltration basin inlet) \\\midrule
\scenario{epsilon} & $67 \unit{km^2}$ highly urban combined sewer network & Maintain sewer network outlet total suspended solids (TSS) load below threshold and avoid flooding (11 in-line storage dams) \\ \midrule
\scenario{zeta} & $1.8 \unit{km^2}$ combined and separated sewer network (based on the Astlingen benchmarking network \citep{Schutze2017, Sun2020}) & Maximize flow to downstream wastewater treatment plant and minimize total combined sewer overflow volume (4 storage basin outlets) \\
\bottomrule
\end{tabular}
\end{table*}
\pystorms abstracts smart stormwater systems into \emph{scenarios}. Each scenario captures a combination of elements that comprise a stormwater control problem. A fully defined scenario includes a network topology (the system or watershed being studied), inputs, a selection of controllable and observable assets, as well as a clearly defined control objective (we refer the reader to Figure~\ref{fig:scenarioComponents} and Table~\ref{tab:terminology} for further details). While users can---and are encouraged to---create their own scenarios, at the time of writing, \pystorms provides an initial collection of seven scenarios, all drawn from real-world smart stormwater systems across North America and Europe. The collection of scenarios spans a variety of stormwater systems that address a diverse set of urban watershed needs with various control objectives. The subcatchment areas range from $0.12 - 67 \unit{km^2}$ in size, and include both combined and separated stormwater arrangements. A summary of the collection of scenarios are presented in Table~\ref{tab:scenarios}, with their more detailed descriptions provided throughout this paper's appendices.  

\

To demonstrate a simple example of a \pystorms scenario, we can focus on Scenario \scenario{theta}, an idealized stormwater network synthesized for unit testing and rapid algorithm exploration. Scenario \scenario{theta}'s \emph{network topology} includes two $1000 \unit{m^3}$ storage basins connected in parallel and draining into a shared downstream water body. The \emph{event driver} is a synthetic rain event lasting $9 \unit{hr}$ with a peak intensity of $3.2 \unit{in} \unit{hr^{-1}}$. The \emph{observable states} are the water levels at the two basins, reported at $15 \unit{min}$ time-steps. The \emph{controllable assets} are valves at the outlet of both storage basins, adjustable at each time-step between $0-100\unit{\%}$ open. The \emph{control objective} is to maintain the outflow into the downstream water body below a specified threshold of $0.5 \unit{m^3s^{-1}}$, while simultaneously preventing flooding at the basins. The ability of a control strategy to meet \scenario{theta}'s control objective is quantified using a pre-defined performance metric that computes a penalty for violating the control objective at each time-step, and sums these penalties across the whole simulation. We provide the specific details on this performance metric (eq.~\ref{eq:perf-obj}) in Section~\ref{sec:evaluatingcontrolstrategies} where we evaluate the performance of two different example control strategies applied to \scenario{theta}.

\

While our aim is for this collection of scenarios to represent a myriad of smart stormwater control applications, we recognize that it is certainly not exhaustive. Ultimately, we aspire to grow the \pystorms repository of stormwater scenarios through community-driven contributions of new scenarios. Accordingly, we provide extensive documentation \footnote{\href{https://klabum.github.io/pystorms/build/html/BuildingScenarios.html}{open-storm.org/pystorms/docs/build-scenarios}} for users to contribute their own scenarios, or modify the existing ones. 
\begin{figure}[ht]
    \centering
	\includegraphics[width=0.9\linewidth]{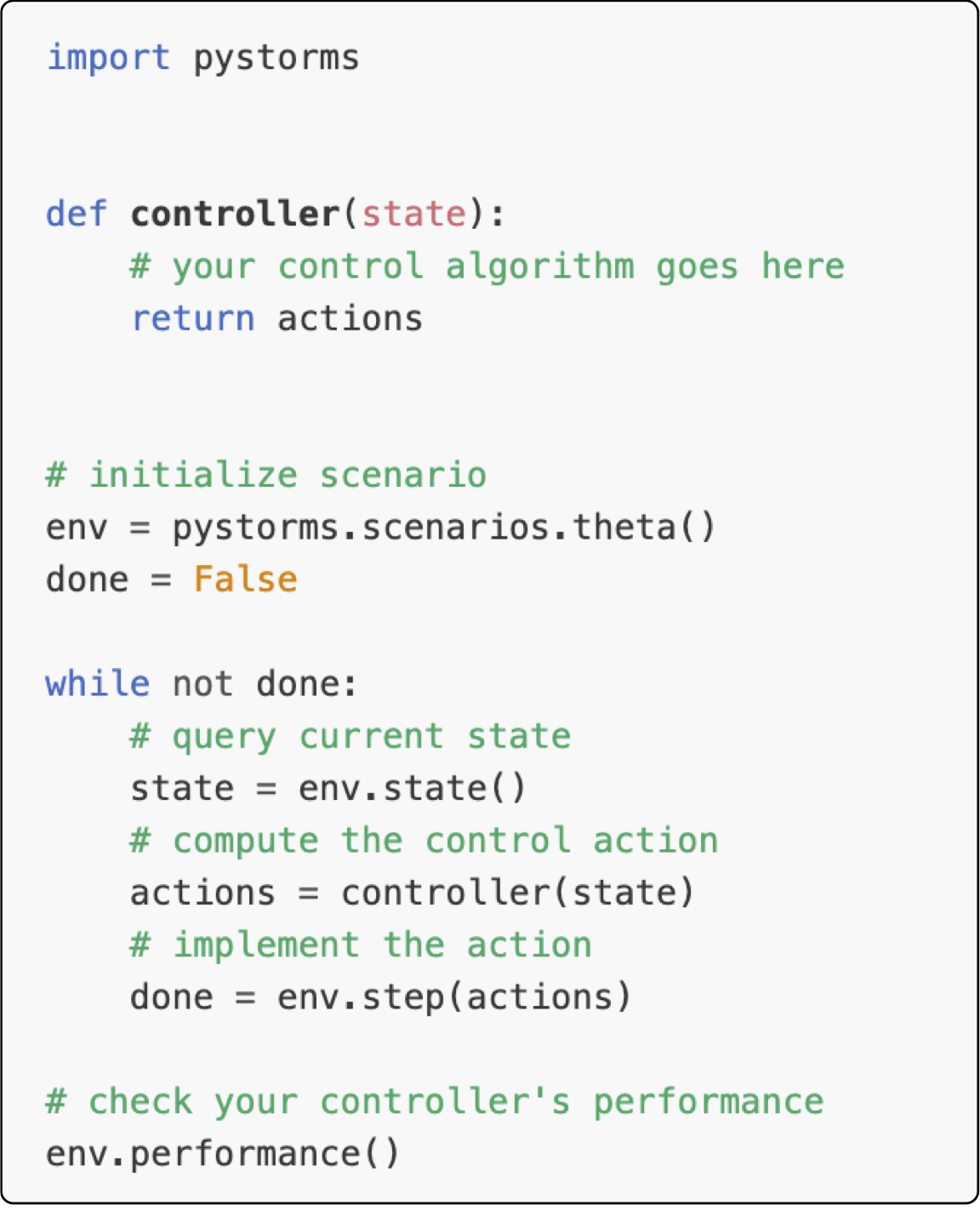}
	\caption{This code snippet is an example implementation of \pystorms for Scenario \scenario{theta}, which is described in greater detail throughout Section \ref{sec:evaluatingcontrolstrategies}. In this example, the controller is implemented as a Python function block.}
	\label{fig:code}
\end{figure}
\subsection{User Experience}
\label{subsec:api}
\pystorms provides a suite of pre-defined smart stormwater systems, and is designed to be both intuitive and accessible for users at any level of expertise or experience of control systems and stormwater management.
Via three intuitive function calls, the user is able to iterate through the simulation of a smart stormwater scenario, and interact with the scenario at any of the simulation timesteps by querying its states or changing the settings of its control assets. 
\pystorms provides the computational environment of a smart stormwater system; the user then provides the control algorithm dictating how it should operate. 
 For developing control strategies, \pystorms allows users the flexibility to utilize any additional computational tools at their disposal. 
 This can be done either by leveraging any additional computational software stacks of Python, or interfacing \pystorms with other computing platforms or languages (e.g., MATLAB, Julia).

\

The user first initializes a \pystorms scenario by creating an instance of it using the statement:  \code{pystorms.scenarios.<scenario name>()}. As seen in the code example (Fig.~\ref{fig:code}), \scenario{theta} can be initialized with \code{pystorms.scenarios.theta()}. The initialization  configures the stormwater simulator with the computational representations necessary to simulate the respective scenario, and returns it as a Python object. This Python object (\code{env} in Fig.~\ref{fig:code}) can be used to progress and/or pause the stormwater simulator, read and/or write parameters to the network, and utilize any additional \pystorms functionality. 

\

The \pystorms programming interface is inspired by the principles of control theory, where the control of a system is abstracted as a control loop in which a controller monitors the underlying state(s) of the system and makes calculated adjustments to the system for it to achieve a desired behavior.  The basic control loop is implemented using the following steps: 
\begin{enumerate}
	\item \label{query} \textbf{Query the set of observable states} in the stormwater network at the current time-step;
	\item \label{compute} \textbf{Compute control actions} to manipulate the system to achieve a desired behavior; and
	 \item \label{implement} \textbf{Implement the control actions} by adjusting the settings of the controllable assets. 
\end{enumerate}

\

 The state of the underlying stormwater network in the scenario can be queried using the \code{<scenario object>.state()} call (see \code{env.state()} in Fig.~\ref{fig:code}).
\code{<scenario object>.step(<actions>)} implements the control actions in the stormwater network, progresses the simulation forward a time-step, and returns the current status of the simulation (\code{True} when the simulation has terminated and \code{False} otherwise). In Fig.~\ref{fig:code}, \code{done = env.step(actions)} implements \code{actions} in the stormwater network and progresses the simulation being handled by the \code{env} Python object, which in this case is the Scenario \scenario{theta} (assigned via \code{pystorms.scenario.theta()}).
\code{done} is assigned \code{True} when the simulation has terminated, and \code{False} otherwise.

\

During the each time-step of the simulation, the metrics that underly the scenario's control objective are evaluated. This computed value is then stored for each time-step, and can be accessed at any time during the simulation using \code{<scenario object>.performance()} (\code{env.performance()} in Fig.~\ref{fig:code}). Additional parameters are logged throughout the simulation. While an initial set of these logged parameters is predefined, the user is able to customize this set for any additional parameters of interest.  Users carry out Steps \ref{query} and \ref{implement} with expressions \code{<scenario object>.state()}, and \\
\code{<scenario object>.step(<actions>)}, respectively. The user defines the controller (Step \ref{compute}), which maps the observed states to control actions. 
\subsection{Architecture}
\label{subsec:architecture}
The \pystorms architecture follows an object oriented programming paradigm and relies on classes as its core building blocks. While the \pystorms programming interface is designed with the intent to be intuitive for all potential users, it particularly caters to those who may only have a rudimentary understanding of stormwater dynamics and/or basic familiarity with programming in Python. However, it can also be customized to meet the requirements of researchers who want to incorporate advanced functionality, such as custom water quality or rainfall-runoff modules. For details on how to utilize \pystorms modularity and customization, we direct the reader to its online documentation.

\

\begin{figure}[ht]
    \centering
	\includegraphics[width=\linewidth]{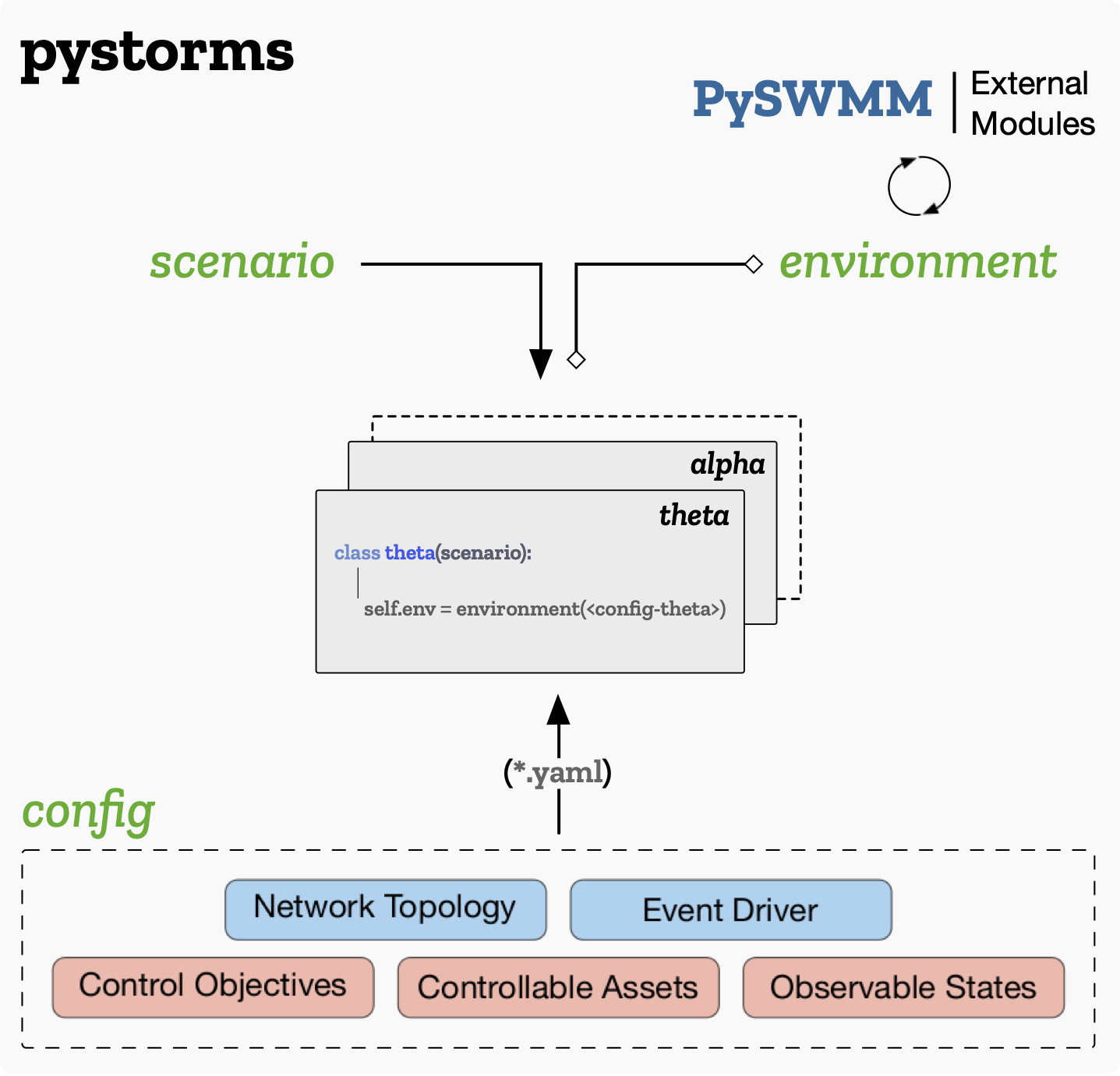}
	\caption{\pystorms is built with three interacting core modules: (i) \texttt{config} represents the metadata and computational representations of the stormwater network and event driver; (ii) \texttt{environment} acts as an interface for scenarios to interact with the stormwater simulators; and (iii) \texttt{scenario} provides a consistent structure for the scenarios in the package. A scenario object in \pystorms inherits (represented by arrows) from the base \texttt{scenario} class, and interfaces (represented by the line) with the stormwater simulator though the \texttt{environment}. }\label{fig:arch}
\end{figure}

The \pystorms architecture is organized to accomplish two tasks: (1) configure the \pystorms scenario, and (2) execute the \pystorms scenario. These two tasks are carried out using three core interacting modules: \texttt{environment}, \texttt{scenario}, and \texttt{config}. These three modules interface with each other to build and execute a given scenario. Fig.~\ref{fig:arch} provides a schematic of this architecture. The first two modules handle the stormwater simulation, while the latter handles the computational representation of the stormwater networks and the metadata pertaining to the control problem (i.e.\ states, actions, and objectives). 
\subsubsection{Configuration}
\label{subsubsec:config}
The \texttt{config} module is used to manage the configuration in the \pystorms architecture. \texttt{config} contains a configuration file for each scenario, which specifies the stormwater network, and then delineates its observable states, controllable assets, and the set of parameters that are used to compute its control objective's corresponding performance metric. The configuration files are written using YAML, a mark-up language commonly used for developing configuration files in software applications. With YAML, the parameters of interest defined in the configuration file are formatted as vertical lists rather than data structures. As a result, the configuration file becomes more human-readable, and creates a simple but scalable workflow for developing scenarios. 
\subsubsection{Simulation}
\label{subsubsec:simulation}
Scenarios in \pystorms are implemented as Python classes. To ensure consistent functionality across scenarios, each scenario is instantiated as its own independent class with an inherited structure from a base \texttt{scenario} module. The scenario classes interface their corresponding configuration files with the stormwater simulator and implement any of the functions specific to that scenario (e.g.\ functions used for computing performance metrics of corresponding control objectives).

\

The \texttt{environment} module is the interface between the stormwater simulator (e.g.\ EPA-SWMM) and the scenarios. This module is specifically included to ensure \pystorms is able to remain agnostic to whatever stormwater simulator is used. For instance, if a user wants to utilize a customized hydrologic solver for simulating stormwater, they can do so by modifying the \texttt{environment} module to call their solver when the scenarios query it, thus ensuring compatibility to a wide array of simulators with minimal overhead. 

\

\pystorms uses SWMM as its default stormwater simulator. SWMM, developed by the U.S. EPA, is an open-source stormwater simulation model that is extensively used for the design and analysis of stormwater systems across the world. SWMM is written in C, a low-level programming language that results in significant computational efficiency. However, the tradeoff for using C becomes SWMM's subsequent difficulty at being interfaced with the latest scientific libraries, which are primarily developed in high-level programming languages, such as Python. As a result, there have been several efforts over the years to build wrappers for SWMM such that its functionality can be exploited via these high-level languages.

\

PySWMM is a Python package that not only provides a wrapper to communicate with SWMM, but also yields a high-level user interface for querying the various stormwater parameters \citep{McDonnell2020}. \pystorms --- by means of the \texttt{environment} module --- interfaces with SWMM using PySWMM, and as a result, all functionality included in PySWMM can also be accessed using \pystormsNOSPACE. Readers are directed to the documentation for additional details and examples to customize \pystorms to meet their requirements. 
\subsection{Software Availability}
\label{subsec:availability}
Developed in Python, \pystorms is supported on all major operating systems (OSX, Windows, and Linux) and can be installed using \texttt{pip}\footnote{\href{https://pypi.org/project/pystorms/}{pypi.org/project/pystorms}}.
\pystorms is distributed under the GNU General Public GPLv3 license\footnote{\href{https://www.gnu.org/licenses/gpl-3.0.html}{gnu.org/licenses/gpl-3.0.html}}, which ensures that this package and its derivatives remain open-source and can be used free of cost. Additionally, source code for the package is available on Github\footnote{\href{https://github.com/kLabUM/pystorms}{github.com/kLabUM/pystorms}} alongside comprehensive documentation and tutorials to utilize and contribute to its broader utilization and development\footnote{\href{https://www.pystorms.org}{pystorms.org}}. 
\begin{figure*}
    \centering
    \includegraphics[width=\linewidth]{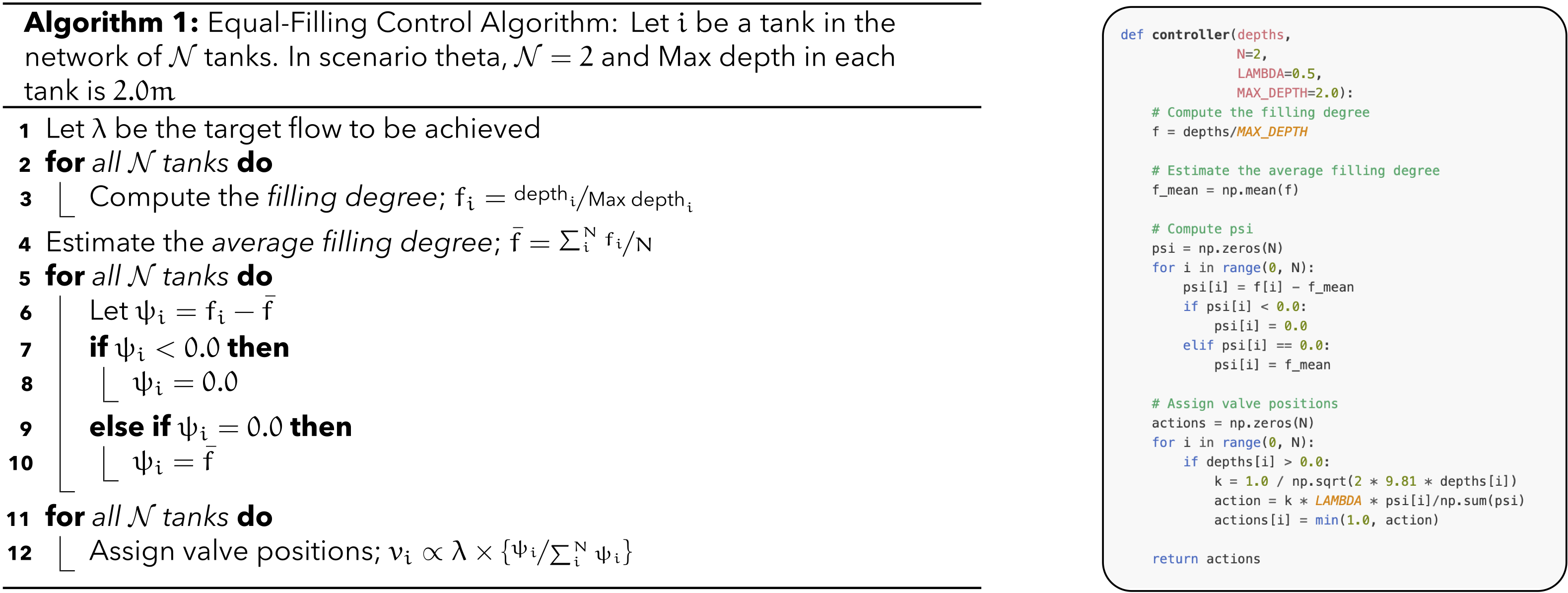}
    \caption{The Equal-filling Controller maintains the flows at the outlet below a desired threshold by coordinating its actions such that it equally utilizes the storage in the controllable assets of the network.
    Algorithm 1 and the corresponding code snippet illustrate the algorithm and its implementation as a function block in Python.    
    An interactive example of the algorithm implementation and its evaluation on Scenario \scenario{theta} can be accessed at \demotheta.}
    \label{fig:eqf}
\end{figure*}
%
%
%
%
%
%
%
%
%
%
%
%
%
%
%
\section{Demo: Implementing and Evaluating Control Strategies}
\label{sec:evaluatingcontrolstrategies}
%
%
%
Here, we demonstrate how \pystorms facilitates developing smart stormwater control strategies by evaluating the performance of two control algorithms applied to Scenario \scenario{theta}.
\subsection{Scenario \scenario{theta}}
\label{subsec:demoscenariotheta}
As stated in Section \ref{subsec:scenarios}, Scenario \scenario{theta} has been developed for rapid prototyping and unit testing of new control strategies. Because Scenario \scenario{theta} is an idealized case, its corresponding performance objective is defined such that a \dquote{perfect} score of $0$ is achievable. We demonstrate how to utilize Scenario \scenario{theta} in this manner by presenting two different control strategies applied.
\subsubsection{Implementating Control Strategies}
\label{subsubsec:demothetaimplement}
While there exist many control strategies that can be adopted to achieve \scenario{theta}'s control objective, we implement and compare two basic strategies here: a simple rule-based control strategy, and the Equal-filling Degree control strategy. Both control strategies adjust the valve openings of \scenario{theta}'s two basin outlets to either retain or release storage depending on the observed states; however, the rule-based control strategy illustrates a simple example that may be explored by a first time user, while the Equal-filling Degree control strategy is presented as an example of an established methodology widely used by stormwater control practitioners. 
\subsubsection*{Rule-based Controller.}
The rule-based control strategy adjusts our basin outlets based on their respective water levels. Specifically, each basin's outlet setting is equal to its relative water level (i.e., the current water level of the basin divided by its maximum depth). Therefore, our control algorithm will set a full basin's outlet to 100\% open, and a basin that is half full will have its outlet set to 50\% open, etc. While this strategy provides a means to mitigate local flooding at each basin, it notably does not consider the other control objective for the network's outflow into the downstream water body to stay below a given threshold.
\subsubsection*{Equal-filling Degree Controller.}
The equal-filling degree is a control strategy often applied to stormwater networks with distributed stormwater storage assets, and has been used in some cases as a starting point when comparing more than one control strategies \citep{Borsanyi2008, Campisano2000, Dirckz2011, Kroll2016, Vezzaro2014}. For this strategy, we begin by defining a storage asset's \dquote{filling degree}---which is typically the ratio a storage asset is full based on its volume or depth---and compute it for each asset in the collection system. The algorithm seeks to \dquote{balance} these filling degrees across the system based on its average. The exact manner in which this balancing is carried out is not necessarily consistent in literature. Our method for this balancing is delineated in the algorithm in Fig.~\ref{fig:eqf}. If all assets have a filling degree equal to the average (i.e., all assets are equally filled), then each should release an equal fraction of the target outflow. Otherwise, the released flows across the assets should be differentiated such that, when an asset has a filling degree less than the average, it does not release any flow; but if an asset is greater than the average, it releases flows based on its deviation from the average. 

\

The implementation of the equal-filling degree algorithm using \pystorms can be seen in Figure~\ref{fig:eqf}. We carry out the simulation for each of the two algorithms, as well as for the \emph{uncontrolled case}, in which control actions are never implemented and the basin outlets are always open. The resulting hydraulic behavior at the two basins and the network's outflow for each of these simulation runs can be seen in Figure~\ref{fig:results}. 
\subsubsection{Evaluating Control Strategies}
\label{subsubsec:demothetaevaluate}
\begin{figure}
    \centering
    \includegraphics[width=\linewidth]{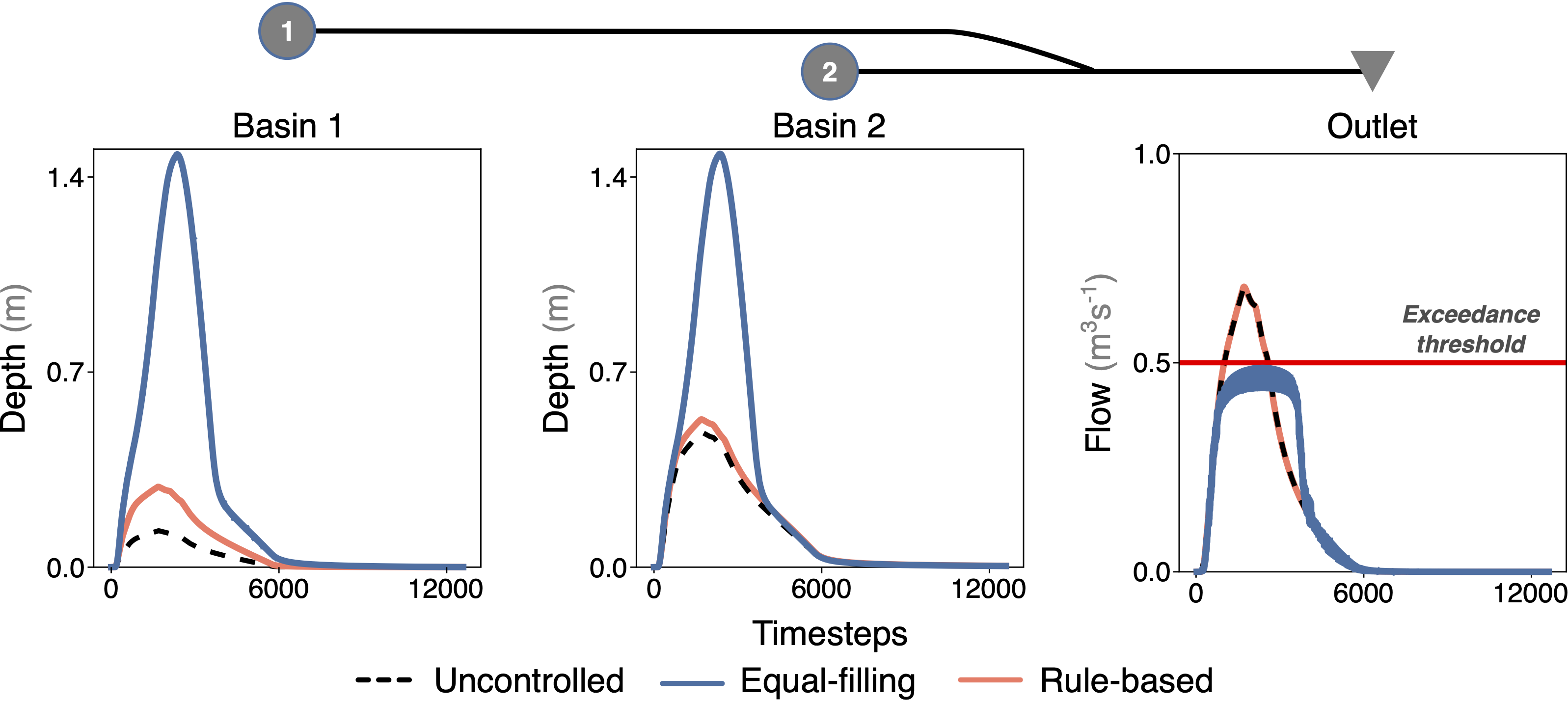}
    \caption{In Scenario \scenario{theta}, the equal-filling degree control strategy is successfully able to maintain the flows at the outlet of the watershed below the desired threshold of $0.5 \unit{m^3s^{-1}}$ by uniformly using the storage in the networks. Static rule-based control and uncontrolled responses of the networks are also presented for comparison. The maximum depth in each of the two basins is $2 \unit{m}$.}
    \label{fig:results}
\end{figure}

The aim of Scenario \scenario{theta} is to find a control strategy that can meet \scenario{theta}'s \emph{control objective} to maintain the outflow into the downstream water body below a specified threshold of $0.5 \unit{m^3s^{-1}}$ and also minimize flooding at the basins. As discussed in Section~\ref{subsec:scenarios}, we pre-define a performance metric to quantify the control algorithm's ability to meet the corresponding control objective. For Scenario \scenario{theta}, this performance metric, $P$, is defined as:
\begin{subequations}\label{eq:perf-sub}
\begin{equation}\label{eq:perf-obj}
	P = \mathlarger{\sum_{t=0}^T} \left( \mathcal{H}_{t} + \displaystyle\sum_{i=1}^2 \mathcal{G}_{i,t}\right)
\end{equation}
\begin{align}
   \mathcal{H}_t &= 
      \begin{cases} 
        Q_{t} - 0.5 \,,
                   & \text{if \,} Q_{t} > 0.5 \\
		   0.0\,,     & \text{otherwise} \\
      \end{cases}
      \label{eq:perfdeviation} \\
   \mathcal{G}_{i,t} &=
      \begin{cases}
        10^3\,,    & \text{if any flooding at basin \,} i \\
        0.0\,,     & \text{otherwise} \\
      \end{cases}
      \label{eq:perfflooding}
\end{align}
\end{subequations}
where $\mathcal{H}_{t}$ is a flow exceedance penalty of the stormwater network's outflow, $Q_{t}$, over the $0.5 \unit{m^3s^{-1}}$ threshold; and $\mathcal{G}_{i,t}$ is an arbitrary flooding penalty of $10^3$ added whenever there exists flooding at either of our two basins, both calculated and summed across every time-step $t$ in the simulation. The performance metric, which is calculated across the simulations for the uncontrolled case and both of the implemented control algorithms, can be seen in Table~\ref{tab:perfmet}. Additionally, corresponding hydraulic behavior for all three cases at their network outlet and both basins can be seen in Fig.~\ref{fig:results}.
The equal-filling degree strategy is able to achieve the control objective of the outflow threshold, as well as avoidance of flooding. Alternatively, the rule-based control strategy only is able to avoid flooding at the basins. The stormwater network behavior for both strategies follow their corresponding implemented algorithm. For example, as the rule-based control strategy does not directly consider the outflow threshold when determining the implemented control actions, it follows that the outflow in the network's outlet exceeds this threshold (see the outlet plot in Fig.~\ref{fig:results}). 

\

The results for each implemented control strategy versus the uncontrolled case are also captured using \scenario{theta}'s performance metric seen in Equation~\ref{eq:perf-sub}. As the performance metric is ultimately a sum of penalties for violating the control objective, a smaller calculated performance metric value indicates a better performing control algorithm. The respective performance metric values for each control strategy presented here can be seen in Table~\ref{tab:perfmet}. With a calculated performance metric of $0$, the equal-filling degree strategy meets \scenario{theta}'s control objective; 
comparatively, the rule-based and uncontrolled cases have higher performance metric values, and thus, we can conclude perform worse than the equal-filling degree.
\begin{table}
\small
\caption{Calculated performance metric values from Equation~\ref{eq:perf-sub} for  simulations corresponding to the two implemented control algorithms and the uncontrolled simulation. As can be seen, the equal-filling degree control strategy performs better than the rule-based control strategy, which then outperforms the uncontrolled case.}\label{tab:perfmet}
\begin{tabular}{l c}
\toprule
{\centering \textbf{Control Strategy}} & \textbf{Performance Metric} \\\midrule
Uncontrolled & 1630 \\
Rule-based & 1624 \\
Equal-filling Degree & 0 \\
\bottomrule
\end{tabular}
\end{table}
%
%
%
\subsection{Additional demos}
\label{subsec:addtldemos}
%
%
%
For demonstration purposes, Scenario \scenario{theta} was chosen due to its simplicity. For more complex examples, we direct the users to the Github repository where a Jupyter Notebook has been developed for each of the scenarios presented in Table \ref{tab:scenarios}, and a control strategy is implemented and evaluated against the uncontrolled case using the corresponding performance metric. The details for each scenario are provided in the \ref{si}. In particular, we direct the reader to Scenarios \scenario{beta} (\ref{si:beta}), \scenario{gamma} (\ref{si:gamma}), \scenario{epsilon} (\ref{si:epsilon}), and \scenario{zeta} (\ref{si:zeta}), in which advanced controllers have been implemented, and the development of these controllers documented in previous research from \citet{Sadler2020},  \citet{Mullapudi2020}, \citet{Troutman2020}, and \citet{Sun2020}, respectively.
%
%
%
%
%
%
%
%
%
%
%
%
\section{Discussion}
\label{sec:discussion}
%
%
%
The ability for stormwater systems to be rapidly modified is critical as communities prepare for climate change and more frequent, uncertain, and destructive weather events. Moreover, \emph{often the most basic control strategies} can have large-scale impacts on the complex, dynamic systems they operate, potentially leading to millions of dollars in savings for the communities they serve. Even though sensor-actuator components may be successfully deployed at individual sites throughout a stormwater network, determining strategies for their coordination across the entire watershed may only add further complexity. As a result, there is a great need --- along with numerous opportunities --- to develop and implement novel control strategies to transform stormwater systems. The sandboxing efforts of \pystorms serves as an initial effort to foster the development of these strategies. Moreover, we intend for \pystorms to serve as a catalyst for our research community to be more expansive and comprehensive in its analysis of smart stormwater control.

\

A critical limitation to progressing smart stormwater control research forward is the inability to systematically develop and then analyze smart stormwater simulation workflows and control strategies. \pystorms directly enables this development and analysis to become more extensive. 
\pystorms can be customized and adapted for a multitude of applications beyond its initially-provided collection of scenarios, and additional research questions can be studied by assembling new scenarios from the assortment of components of this scenario collection. For example, for each of the scenarios we provide at the outset, \pystorms specifies only a subset of the scenario's total observable states that are able to be queried throughout the simulation.
But this initial subset of observable states is never claimed as the optimal; and actually, identifying an optimal set of observable states is in and of itself a rich area of research \citep{ChaconHurtado2017, Sambito2019, Bartos2020, VanNguyen2020}. Thus, new scenarios can be made with different subsets of observable states (e.g.\ flows, pollutant concentrations), and new research questions can now be asked about which states may be most critical for informing control actions to be taken.

\

In the current iteration of \pystormsNOSPACE, we present one means for assessing a control strategy via the scenario's corresponding performance metric (e.g. maintain flow below a threshold, avoid flooding). 
However, \pystorms can serve as a mechanism for developing and implementing additional metrics. For example, by increasing the number of controllable assets available out of the eleven pond outlets presented in Scenario \scenario{gamma}, one can assess the \emph{scalability} of a control algorithm as the state-action space increases. Additionally, control algorithm \emph{generalizability} across storm characteristics can be assessed with the multiple rain events provided in Scenario \scenario{epsilon}. 

\

Beyond the coordination and integration of smart stormwater control methods, we also view a more expansive --- and potentially consequential --- opportunity for \pystorms to drive our research community's analysis of \dquote{control} to include the social and ethical implications of its implementation. 
We recognize that the control of stormwater systems operates within broader, and far more complex, socio-techinical systems, and as a result, the decisions made can have a profound impact on the lives of people within those systems.
Thus, in addition to water quantity and quality analyses, we impel our community to also scrutinize their control strategies through the lens of social equity and longterm community resilience and adaptiveness, such as the decision framework developed by \citet{Ewing2021}. We believe \pystorms can help facilitate these analyses. 
For example, for any of the provided scenarios, one could study the longterm social implications of an adverse outcome that might occur regularly in the same community due to an implemented control strategy by modifying a scenario's control objective and corresponding performance metric.   

\

Finally, we encourage future users of \pystorms to resist the natural inclination towards fixating on a control algorithm's \emph{performance} at the expense of its \emph{appropriateness}. Our development of \pystorms was partly motivated by similar sets of data and libraries developed by the electrical and computer science communities over the past two decades in an effort to evaluate their ever-growing assortment of machine-learning algorithms \citep{Deng2012, Russakovsky2015, Brockman2016, Henderson2017}. While their efforts have been wildly successful at propelling the development of novel algorithms, it has also been noted that much of that effort has been expended in the \dquote{fine-tuning} and \dquote{hacking} of the various algorithms to achieve some sort of arbitrarily-defined \dquote{state-of-the-art} result \citep{Torralba2011}. While fine-tuning an algorithm's behavior for a specific dataset may indeed yield a better performance metric in \pystormsNOSPACE, this improved performance metric is not guaranteed to translate to the physical system at hand. Validating these stormwater control strategies requires much further analysis via their actual physical deployment. Accordingly, we intend for \pystorms to foster a collaborative and thoughtful evaluation of stormwater control algorithms rather than a myopic focus on performance metrics and competition.
%
%
%
%
%
%
%
%
%
%
%
%
%
%
%
\section{Conclusions and Next Steps}
\label{sec:conclusions}
%
%
%
\pystorms provides a curated collection of scenarios, coupled with an accessible programming interface, to enable the development and quantitative evaluation of stormwater control algorithms. 
We have developed \pystorms with the intent to make smart stormwater control research more methodical and efficient.  
As shown with the demos in Section \ref{sec:evaluatingcontrolstrategies} and accompanying tutorials, we have demonstrated how users can quickly download and test \pystorms and its scenarios on a basic computer with only a few lines of code.
Additionally shown in Section \ref{sec:evaluatingcontrolstrategies}, \pystorms facilitates rigorous evaluation by extricating the control algorithm implementation of a stormwater control strategy to be applied and quantitatively compared across the example scenarios.

\

It is our hope that this package will emerge as a community-driven resource that is able to address key knowledge gaps and enable the advancement of smart stormwater systems. 
To this extent, we see proximate opportunities for the broader research community to collaborate on \pystorms by contributing their own stormwater scenarios and/or control algorithms to the package initiated here. 
Likewise, we encourage the broader research community to further build upon \pystorms by imparting their own smart stormwater control instances using the \pystorms architecture and integrating their own stormwater control simulation workflows into it.
%
%
%
%
%
%
%
%
%
%
%
%
%
%
%
\begin{acks}
This research was supported U.S. National Science Foundation, Award Numbers: 1737432 and 1750744.
Additionally, Argonne National Laboratory's contribution is based upon work supported by Laboratory Directed Research and Development (LDRD) funding from Argonne National Laboratory, provided by the Director, Office of Science, of the U.S. Department of Energy under Contract No. DE-AC02-06CH11357.
\end{acks}
%
%
%
%
%
%
%
%
%
%
%
%
%
%
%

%
%
%
%
%
\onecolumn
\newpage

\newcommand{\beginsupplement}{
        \setcounter{table}{0}
        \renewcommand{\thetable}{S\arabic{table}}
        \setcounter{figure}{0}
        \renewcommand{\thefigure}{S\arabic{figure}}
        \setcounter{section}{0}
        \renewcommand{\thesection}{Supplementary Information}
        \setcounter{subsection}{0}
        \renewcommand{\thesubsection}{SI-\arabic{subsection}.}
     }

\beginsupplement
\section{}
\label{si}
\subsection{Scenario \scenario{alpha}}
\label{si:alpha}
\begin{figure}[ht!]
    \centering
    \includegraphics[width=0.80\textwidth]{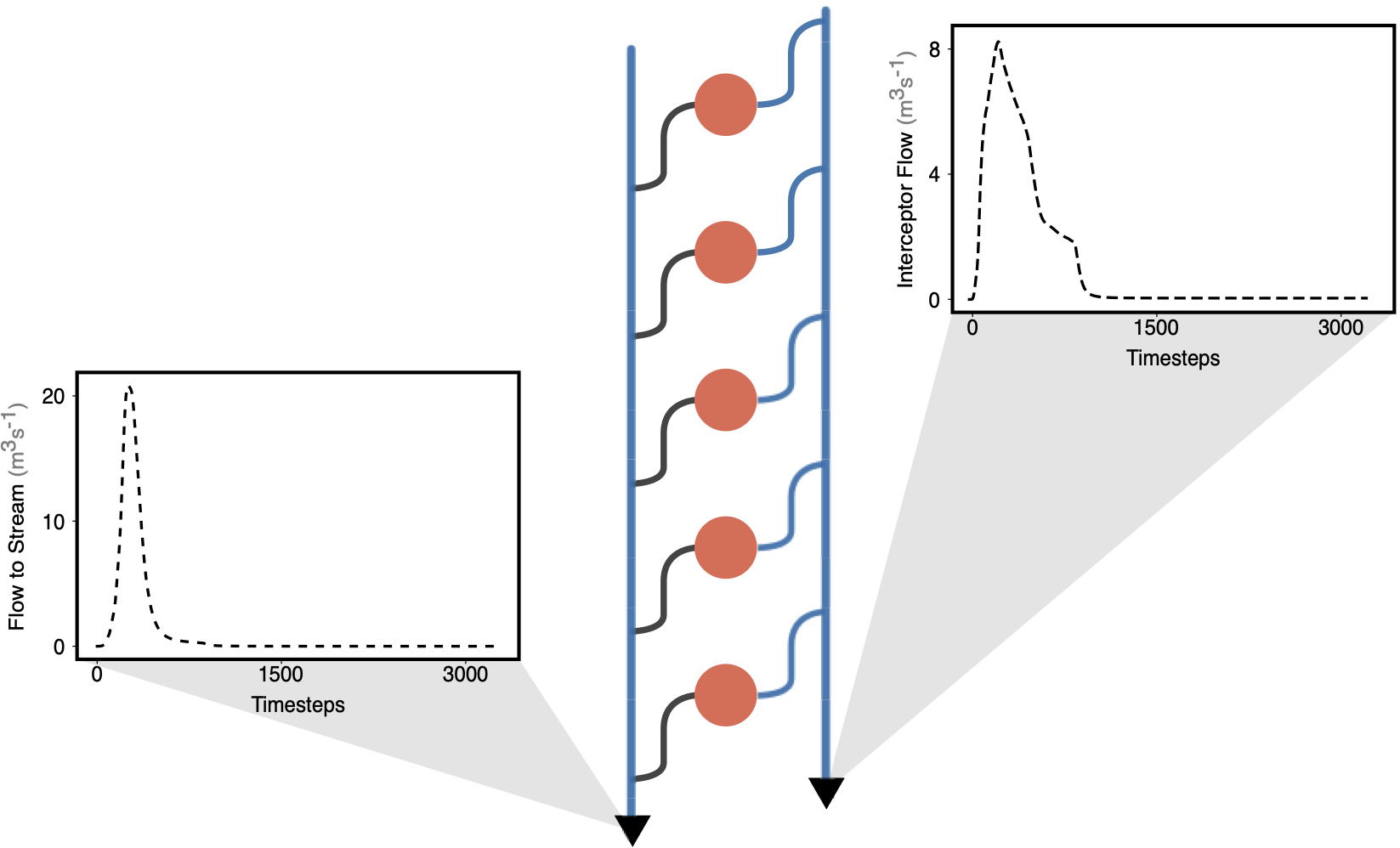}
    \caption{Topology of Scenario \scenario{alpha}'s stormwater network. Regulators control flow of combined stormwater/dry-weather flows to WWTP, and to stream when capacity to treatment plant is exceeded. Plot shows flowrates at downstream locations of stream and interceptor for uncontrolled case.}\label{fig:network_alpha}
\end{figure}

\begin{table}[ht]
\small
\caption{Scenario \scenario{alpha} Configuration}
\begin{tabular}{l p{5.5in}}
\toprule
\textbf{Network Topology} & 
Scenario \scenario{alpha} is a combined-sewer overflow system in a $0.12$ $\unit{km}^2$ residential area, and is based on the SWMM Applications Manual, Example 8 \citep{Gironas2010}. This specific network is chosen to demonstrate how combined sewer overflow (CSO) regulators, which are static in the SWMM Applications' interpretation, can become \textit{active} (e.g. valves whose openings can be remotely controlled), and their coordination can then potentially improve performance. 
\vspace{0.5em}
\newline
Scenario \scenario{alpha} has a series of interceptor pipes that are able to capture and convey 100\% of sanitary flows from sewers to a WWTP during dry weather events. During wet weather events, the capacity of the interceptor pipes may be exceeded. Thus, the sewers are also connected to a nearby water body in order to discharge the excess wastewater during such events (i.e. CSO events). 
\vspace{0.5em}
\newline
To control the flow between the sewer system, the interceptor pipes, and the water body, a set of five flow regulators are implemented in parallel to one another. The five flow regulators are implemented as transverse weirs with orifices, in which the sanitary and stormwater flows are conveyed to the interceptor through a controllable orifice, and excess flow is then diverted to the overflow outlet into the body of water via a weir upstream of the orifice. 
\vspace{0.5em}
\newline
The network topology for Scenario \scenario{alpha} is in Figure \ref{fig:network_alpha}.
\\\midrule
\textbf{Event Driver} & 
Scenario \scenario{alpha} is driven by a 10-year, 2-hour storm event. It is assumed the entire subcatchment experiences uniform rainfall. \\\midrule
\textbf{Controllable Assets} & 
The controllable assets of Scenario \scenario{alpha} are a set of five orifice valves that control the wastewater flow to the interceptor pipes. The valve openings are controllable between 0\% -- 100\%, where the percentage of each orifice valve corresponds to the percent opening for the wastewater flow into the interceptor pipes; all orifice valves are initially set to 100\% open. 
\vspace{0.5em}
\newline
The orifice valves are one component of a flow regulator. For each flow regulator, once the flow capacity through its orifice valve is reached, the remaining wastewater is diverted upstream, potentially overtopping a transverse weir, and then \emph{overflowing} into the nearby stream via the corresponding CSO outfall.  
\vspace{0.5em}
\newline
In the computational implementation of Scenario \scenario{alpha}, the controllable valves for each of the five regulators are identified with the pattern \texttt{<Or1, Or2, Or3, Or4, Or5>}.\\\midrule
\textbf{Observable States} & 
The observable states for Scenario \scenario{alpha} include the following:
\begin{itemize}
\item Depth at each runoff entry point into sewer network
\item Depth if the junctions nodes at the sewer inlet into the flow regulators
\item Depth at the CSO outfalls
\item Depth at each segment of the stream
\item Flowrate at final interceptor pipe connected to WWTP
\end{itemize}
The values for each can be queried throughout the simulation to inform any control decisions. \\\midrule
\textbf{Control Objectives} & 
The primary objective for Scenario alpha is to minimize the total CSO volume across the system, while also avoiding flooding. The ability to achieve the objective of Scenario \scenario{alpha} is quantified via its performance metric, $P$, in the equation below.
\begin{center}
\begin{math}
\begin{aligned}[t]
P = \displaystyle\sum_{t=0}^T\left[\displaystyle\sum_{i=1}^5 V_{\mathrm{CSO}_i}(t) + \left(10^6 \cdot\displaystyle\sum_{j=1}^{26} V_{\mathrm{flood}_j}(t)\right) \right]
\end{aligned}
\end{math}
\end{center}
where $V_{\mathrm{CSO}_i}(t)$ is the volume of CSO overflow at regulator $i$ over timestep $t$, and $V_{\mathrm{flood}_j}(t)$ is the volume of flooding over timestep $t$ at all of the 26 junction nodes in the network. (Note: the flooding volume is purposely penalized by a factor of $10^6$.)
\\ \midrule
\textbf{Additional Notes} & 
As stated previously, Scenario \scenario{alpha} is based on the eighth example model presented in the SWMM 5.0 Applications Manual \citep{Gironas2010}, which was developed to demonstrate how to represent a combined sewer system in SWMM, and included flow regulators as well as a downstream pump station as control assets. In the SWMM Applications Manual, the outflow of the interceptor leads to a storage well that is then pumped through a forcemain pipe to the WWTP, and a rule-based controller dictates when a pump is operated based on depth of the well. This forcemain pipe and corresponding pump station are not included in Scenario \scenario{beta}. However, these components could be added as an additional component to this network for further dynamic control testing and exploration.
Interactive example: \demoalpha\\
\bottomrule
\end{tabular}
\end{table}
\clearpage
\subsection{Scenario \scenario{beta}}
\label{si:beta}
\begin{figure}[ht!]
    \centering
    \includegraphics[width=0.70\textwidth]{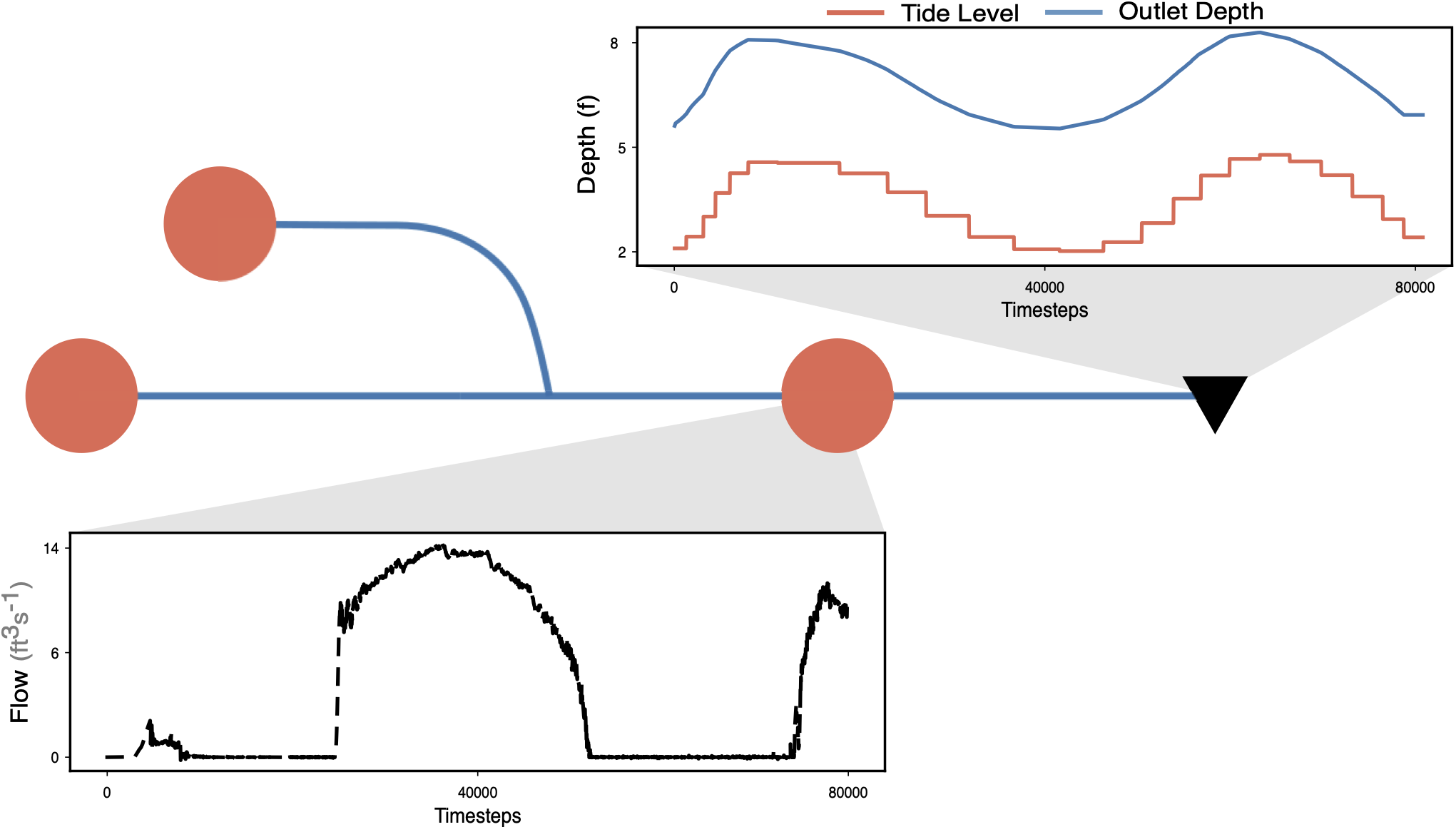}
    \caption{The network schematic for Scenario \scenario{beta}. Orange circles and blue lines in the figure represent, respectively, the controllable storage elements and channels draining into the outlet (black triangle). 
    Scenario \scenario{beta} has three control assets: furthest upstream is a valve at the outlet of a stormwater pond for controlling the pond's storage, connected in parallel to a pump used to increase the hydraulic head of the underlying stormwater pipes, which both drain into shared downstream conduit with inline storage controlled by an inflatable dam.
    To demonstrate the tidal fluctuation in this scenario, the flowrate at the node with the lowest elevation (i.e. junction \texttt{J33}) is plotted against the depth at the network outfall \texttt{OUT0}, which has an assumed increase in sea level of $3 \unit{ft}$.}\label{fig:network_beta}
\end{figure}
\begin{table}[ht]
\small
\caption{Scenario \scenario{beta} Configuration}
\begin{tabular}{l p{5.5in}}
\toprule
\textbf{Network Topology} & 
Scenario \scenario{beta} is a separated storm system in a highly-urbanized area that drains to a tidally influenced river. The drainage area is $\approx 1.3\ km^2$. The elevation of the system is low and the outfall of the system becomes inundated with certain high tide conditions, making the drainage of the water through the outfall impossible. The system has an upstream stormwater pond, a pump that can help reduce flooding within the system, and variable downstream inline storage via an inflatable dam. Given the constraints on the system, totally eliminating flooding is not always possible. The stormwater network topology for Scenario \scenario{beta} is shown in Figure \ref{fig:network_beta}. \\\midrule
\textbf{Event Driver} & 
Scenario \scenario{beta} is driven by a 2-year, 12-hour storm event with a SCS Type II temporal distribution. 
Scenario \scenario{beta} also has predefined tidal conditions, which are based on a typical tide cycle from a nearby tide gauge. The tidal conditions were adjusted such that high tides equaled between the $99\%$ and $50\%$ annual tide level for the station.
Finally, to test potential future sea level rise scenarios, Scenario \scenario{beta} also assumes a high sea level rise of $3.0 \unit{ft}$, which is incorporated into the scenario via the pre-defined tidal conditions.
\\\midrule
\textbf{Controllable Assets} & 
Scenario \scenario{beta} has three controllable assets: (i) a valve at the outlet of a stormwater pond for controlling the pond's storage, (ii) a pump to increase the hydraulic head of the underlying stormwater pipes, and (iii) an inflatable dam for utilizing inline storage of stormwater pipes. The valve is computationally implemented as an orifice, identified as \texttt{O2} and adjustable to a setting within a range of $0.0$ - $1.0$ (i.e. 0\% to 100\% open). The pump, identified as \texttt{P0}, is computationally implemented with a predefined pump flowrate curve, and can be set to either $0.0$ (i.e., \texttt{OFF}) or $1.0$ (i.e., \texttt{ON}). The inline storage, identified as \texttt{W0}, is computationally implemented as a weir and adjustable to a setting within a range of $0.0$ - $1.0$ (i.e. 0\% to 100\% open). The settings of the three control assets are initially set to $[1.0, 0.0, 1.0]$, which corresponds to a completely open orifice, a pump set to \texttt{OFF}, and a completely open weir.
\\\midrule
\textbf{Observable States} & 
The states in Scenario \scenario{beta} are observable at seven different locations, including \texttt{<}\texttt{J33}, \texttt{J64}, \texttt{J98}, \texttt{J102}, \texttt{OUT0}, \texttt{ST0}, \texttt{ST2}\texttt{>}. 
\\\midrule
\textbf{Control Objectives} & 
The objective for Scenario \scenario{beta} is to minimize flooding at the most critical nodes within the system. This is quantified in the below performance equation.
\begin{center}
\begin{math}
\begin{aligned}[t]
P = \displaystyle\sum_{i=1}^{12}\displaystyle\sum_{t=0}^{T} V_{\mathrm{flood}_i}(t)
\end{aligned}
\end{math}
\end{center}
where $V_{\mathrm{flood}_i}$ is the flooding volume (in gallons) over timestep $t$ across a set of $i=12$ critical nodes. The set of critical nodes is: \texttt{<}\texttt{J4}, \texttt{J8}, \texttt{J13}, \texttt{J33}, \texttt{J53}, \texttt{J54}, \texttt{J64}, \texttt{J65}, \texttt{J98}, \texttt{J102}, \texttt{J145}, \texttt{J146}\texttt{>}.
\\ \midrule
\textbf{Additional Notes} & 
Scenario \scenario{beta} was developed based on a stormwater model developed and tested by \citet{Sadler2020}, with the model, data, and the implementation of its proposed control strategy shared via the HydroShare collaborative website and available at \href{https://doi.org/10.4211/hs.5148675c6a5841e686a3b6aec67a38ee}{https://doi.org/10.4211/hs.5148675c6a5841e686a3b6aec67a38ee}. 
\vspace{0.5em}
\newline
A notebook demonstrating the implementation of Scenario \scenario{beta} is available here: \demobeta.
\\
\bottomrule
\end{tabular}
\end{table}
\clearpage
\subsection{Scenario \scenario{gamma}}
\label{si:gamma}
\begin{figure}[ht!]
    \centering
    \includegraphics[width=0.80\textwidth]{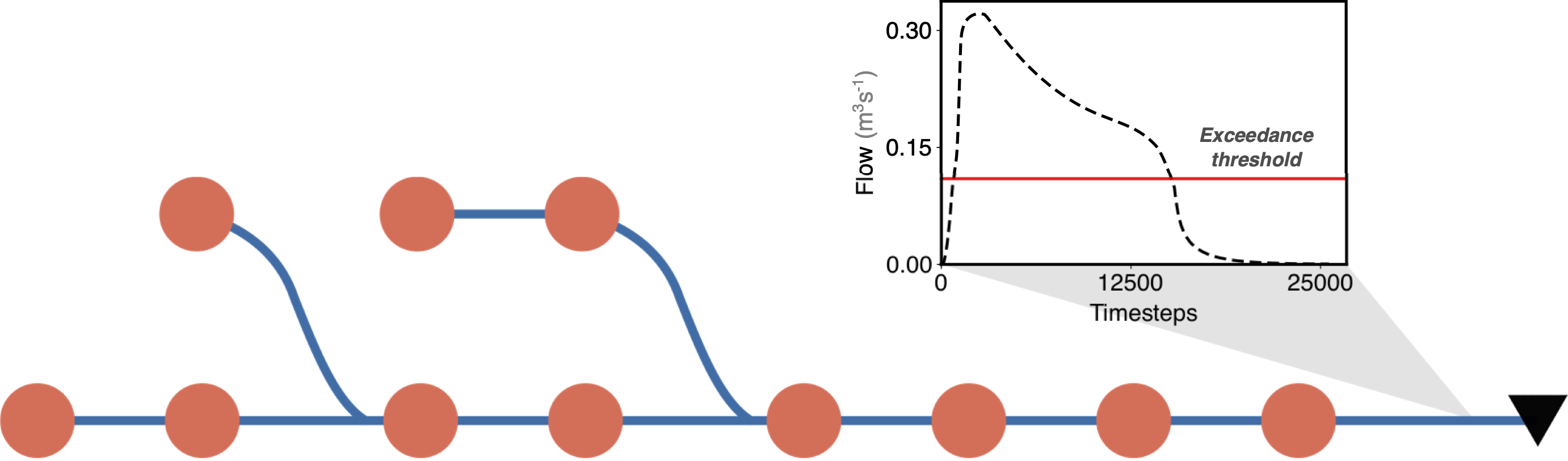}
    \caption{Topology for the separated stormwater network of Scenario \scenario{gamma}.
	Orange circles and blue lines in the figure represent respectively the controllable storage elements and channels draining into the outlet (black triangle). 
	During a storm event, each node in the network receives runoff generated by connected subcatchments in addition to discharge from upstream nodes. 
	The system's response to the given storm event for the uncontrolled case (i.e. all 11 of its valves set to 100\% open) is depicted via plotting the outlet's flows against the target exceedance threshold for each timestep of the simulation. 
	For this scenario, the controller is penalized for any flowrates above given exceedance thresholds for each of the network's 11 links situated immediately downstream from a corresponding control asset.}\label{fig:network_gamma}
\end{figure}
\begin{table}[ht]
\small
\caption{Scenario \scenario{gamma} Configuration}
\begin{tabular}{l p{5.5in}}
\toprule
\textbf{Network Topology} & 
Scenario \scenario{gamma} is a separated system in a semi-urbanized watershed of $\approx 4\ km^2$. 
Scenario \scenario{gamma}'s stormwater network is comprised of 11 storage basins, connected in both series and parallel, that drain into a shared downstream water-body. The stormwater network topology for Scenario \scenario{gamma} can be seen in Figure \ref{fig:network_gamma}. \\\midrule
\textbf{Event Driver} & 
Scenario \scenario{gamma} is driven by a 25-year, 6-hour storm event. 
It is assumed the entire subcatchment experiences uniform rainfall. \\\midrule
\textbf{Controllable Assets} & 
Each of Scenario \scenario{gamma}'s storage basins are control assets equipped with valves adjustable to a setting within a range of (0\% to 100\%) during the simulation. 
The percentage of each valve corresponds to the percent opening of the corresponding storage basin's outlet, with all valves initially set to 100\% open. 
In the computational implementation of Scenario \scenario{gamma}, each of the controllable valves are identified with the pattern \texttt{O<storage asset ID>} (e.g. O1, O2, \ldots O11).\\\midrule
\textbf{Observable States} & 
The observable states for Scenario \scenario{gamma} are the water levels in the the storage basins, which can be queried throughout the simulation to inform any control decisions. \\\midrule
\textbf{Control Objectives} & 
The primary objective for Scenario \scenario{gamma} is to maintain flow in each connecting channel below a given threshold ($0.11\ m^3/s$) while also avoiding flooding. The ability to achieve Scenario \scenario{gamma}'s objective is quantified via its performance, $P$, in the equation below. As seen from the equation, flooding constitutes a much greater penalty than threshold exceedance. Thus, it is presumed a control algorithm will permit threshold exceedance when necessary to avoid flooding. 
\begin{math}
\begin{aligned}[t]
P_i(t) &= \begin{cases}
0.0, & Q_i(t) \leq 0.11 \quad and \quad  F_i(t) = 0 \\
Q_i(t) - 0.11, & Q_i(t) > 0.11 \quad and \quad  F_i(t) = 0 \\
10^6, & F_i(t) > 0
\end{cases}\\
P &= \sum^{11}_{i=1} \sum^T_{t=0} P_i(t)
\end{aligned}
\end{math}
where $Q_i(t)$ and  $F_i(t)$ represent the outflow and flooding, respectively, at the $i^{th}$ node of the network for time $t$.
\\ \midrule
\textbf{Additional Notes} & Interactive example: \demogamma\\
\bottomrule
\end{tabular}
\end{table}
\clearpage
\subsection{Scenario \scenario{delta}}
\label{si:delta}
\begin{figure}[ht!]
    \centering
    \includegraphics[width=0.90\textwidth]{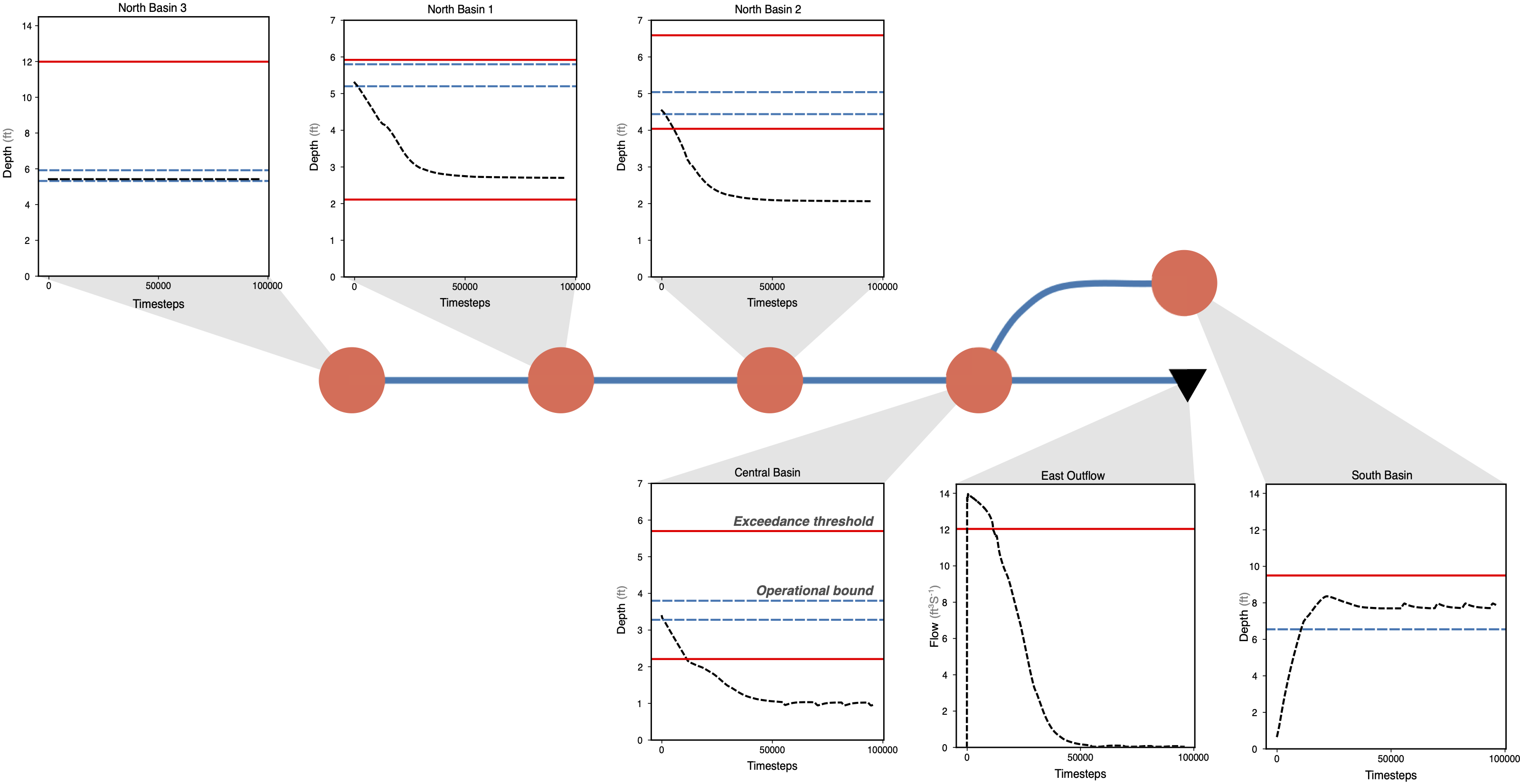}
    \caption{Topology for the residential network of Scenario \scenario{delta}.
    Orange circles and blue lines in the figure represent, respectively, the controllable storage elements and channels draining the outflow (black triangle) of the subnetwork. 
	During a storm event, each storage element in the network receives runoff generated by connected subcatchments, in addition to discharge from upstream storage elements. 
	The system's response to the given storm event for the uncontrolled case (i.e. all 5 of its valves set to 100\% open) is depicted  for each timestep of the simulation by plotting the depths of the basins against their operational bounds and exceedence thresholds, as well as the subnetwork's outflow against its exceedance threshold. 
	For this scenario, the controller is penalized for any depths and flows outside of their thresholds, as well as for any network flooding.}\label{fig:network_delta}
\end{figure}
\begin{table}[ht]
\small
\caption{Scenario \scenario{delta} Configuration}
\begin{tabular}{l p{5.5in}}
\toprule
\textbf{Network Topology} & 
Scenario \scenario{delta} is a subnetwork of a broader combined urban stormwater network. It is a in a residential neighborhood covering approximately $\approx 2.5\ km^2$. 
Scenario \scenario{delta}'s stormwater network is comprised of 6 storage basins: 5 basins connected in series, 4 of which have controllable outlet valves (North Basin 1, North Basin 2, North Basin 3, and Central Basin), and the furthest downstream storage basin (Central Basin) draining into the broader stormwater network through the outfall conduit. Additionally, the Central Basin connects to an infiltration basin (South Basin); the South Basic is equipped with a valve controlling its inflow from the Central Basin. 
\vspace{0.5em}
\newline
While all of these basins are part of the broader network's stormwater control, five of them also serve as waterfront to homes in the community (North Basin 1, North Basin 2, North Basin 3, North Basin 4, and Central Basin). Thus, it is desired to maintain the water level at these basins within a pre-specified range. (Note: As North Basin 4 drains directly into North Basin 3 without an active control mechanism, we are only concerned with monitoring the water level in North Basin 3.)
\vspace{0.5em}
\newline
The stormwater network topology for Scenario \scenario{delta} can be seen in Figure \ref{fig:network_delta}. \\\midrule
\textbf{Event Driver} & 
Scenario \scenario{delta} is driven by an observed storm event (13-hr, 0.93-in) that has an approximate return period of 2-months. It is assumed the entire subcatchment experiences uniform rainfall. Additionally, the conduit connecting the subnetwork to the broader stormwater system experiences constant wastewater flow.  
\\\midrule
\textbf{Controllable Assets} & 
Five of Scenario \scenario{delta}'s storage basins are control assets equipped with valves adjustable to a setting within a range of (0\% to 100\%) during the simulation. The South Basin is an infiltration basin, thus, the valve controls flow \textit{into} the basin. Otherwise, the valves of the other four basins (i.e. North Basin 1, North Basin 2, North Basin 3, and Central Basin) control the flow \textit{out} of each basin. Each basin's valve is initially set to 100\% open. The basin furthest upstream (i.e. North Basin 4) is not controllable and instead flows directly into North Basin 3.
In the computational implementation of Scenario \scenario{delta}, each of the controllable valves are identified as \texttt{weir\_N3}, \texttt{weir\_N2}, \texttt{weir\_N1}, \texttt{weir\_C}, and \texttt{orifice\_S}. 
\\\midrule
\textbf{Observable States} & 
The observable states for Scenario \scenario{delta} are the water levels at the storage basins with controllable valves (i.e., \texttt{basin\_N1}, \texttt{basin\_N2}, \texttt{basin\_N3}, \texttt{basin\_C}, and \texttt{basin\_S}), and can be queried throughout the simulation to inform any control decisions. \\\midrule
\textbf{Control Objectives} & 
The primary objective for Scenario \scenario{delta} is to maintain the depths at the five storage ponds within corresponding inner thresholds. Additionally, any control strategy should avoid flooding and maintain a network outflow below a flowrate threshold. The performance is quantified during each time step of the simulation with penalties increasing exponentially the greater the depth exceeds its predefined operational bounds, and a maximum penalty for depth's at basin's passing their given exceedence thresholds. The ability to achieve Scenario \scenario{delta}'s objective is quantified via its performance, $P$, in the following equation: 
\begin{center}
\begin{math}
\begin{aligned}[t]
P = \sum^T_{t=0} \left(\sum^{5}_{i=1} P_i(t) + \sum^{9}_{j=1} P_j(t) +  P_{\mathrm{out}}(t)\right) 
\end{aligned}
\end{math}
\end{center}
with penalties for the depth at each of the basins computed with
\begin{center}
\begin{math}
\begin{aligned}[t]
P_i(t) &= 
\begin{dcases}
0.0, & O_l \leq d_i \leq O_u \\
\left[\left(10^6 + 1\right)^{\left(\frac{O_l - d_i}{O_l - E_l}\right)}\right] - 1, &  E_l \leq d_i \leq O_l \\
\left[\left(10^6 + 1\right)^{\left(\frac{d_i-O_u}{E_u-O_u}\right)}\right] - 1, &  O_u \leq d_i \leq E_u  \\
10^6, & d_i < E_l \quad or \quad d_i > E_u
\end{dcases} 
\end{aligned}
\end{math}
\end{center}
for each basin $i$ in \texttt{<}\texttt{basin\_N3}, \texttt{basin\_N2}, \texttt{basin\_N1}, \texttt{basin\_C}, \texttt{basin\_S}\texttt{>}, and where $O$ is the corresponding operational bound and $E$ is the exceedence threshold defined for each basin; penalties from flooding 
computed as
\begin{center}
\begin{math}
\begin{aligned}[t]
P_j(t) &= 
\begin{dcases}
0.0, & F_j = 0  \\
10^6, & F_j > 0
\end{dcases} 
\end{aligned}
\end{math}
\end{center}
for $j$ in the following set of network nodes:  \texttt{<}\texttt{junc\_N4sc}, \texttt{basin\_N4}, \texttt{junc\_N3sc}, \texttt{basin\_N3}, \texttt{junc\_N2sc}, \texttt{basin\_N2}, \texttt{junc\_N1sc}, \texttt{basin\_N1}, \texttt{junc\_Csc}, \texttt{basin\_C}, \texttt{junc\_Ssc}, \texttt{basin\_S}, \texttt{junc\_EinflowA} \texttt{>}; and penalties for exceeding the outflow threshold computed as
\begin{center}
\begin{math}
\begin{aligned}[t]
P_{\mathrm{out}}(t) = 
\begin{dcases}
0.0, & Q_{\mathrm{out}} \leq 12.0  \\
\left(1 - \dfrac{12.0}{Q_{\mathrm{out}}} \right)^6, &  Q_{\mathrm{out}} > 12.0
\end{dcases} 
\end{aligned}
\end{math}
\end{center}
at the network's outflow \texttt{conduit\_Eout}. The upper and lower operational bounds and exceedence thresholds for each basin are delineated in the following table (Note: the South Basin does not have a lower bound/threshold):
\begin{center}
\begin{tabular}{l | p{0.75in} | p{0.75in} | p{0.75in} | p{0.75in} |}
& \multicolumn{2}{c|}{Operational bound (O)} & \multicolumn{2}{c|}{Exceedence threshold (E)} \\
& Lower limit, $O_l$ & Upper limit, $O_u$ & Lower limit, $E_l$ & Upper limit, $E_u$ \\\hline
North Basin 3 & \multicolumn{1}{|c|}{$5.32$} & \multicolumn{1}{|c|}{$5.92$} & \multicolumn{1}{|c|}{$5.28$} & \multicolumn{1}{|c|}{$11.99$} \\ 
North Basin 2 & \multicolumn{1}{|c|}{$4.44$} & \multicolumn{1}{|c|}{$5.04$} & \multicolumn{1}{|c|}{$4.04$} & \multicolumn{1}{|c|}{$6.59$} \\ 
North Basin 1 & \multicolumn{1}{|c|}{$5.20$} & \multicolumn{1}{|c|}{$5.80$} & \multicolumn{1}{|c|}{$2.11$} & \multicolumn{1}{|c|}{$5.92$} \\ 
Central Basin & \multicolumn{1}{|c|}{$3.28$} & \multicolumn{1}{|c|}{$3.80$} & \multicolumn{1}{|c|}{$2.21$} & \multicolumn{1}{|c|}{$5.70$} \\ 
South Basin	& \multicolumn{1}{|c|}{---} & \multicolumn{1}{|c|}{$6.55$} & \multicolumn{1}{|c|}{---} & \multicolumn{1}{|c|}{$9.55$} \\ 
\end{tabular}
\end{center}
\\ \midrule
\textbf{Additional Notes} & Interactive example: \demodelta\\
\bottomrule
\end{tabular}
\end{table}
\clearpage
\subsection{Scenario \scenario{epsilon}}
\label{si:epsilon}
\begin{figure}[ht!]
    \centering
    \includegraphics[width=0.7\textwidth]{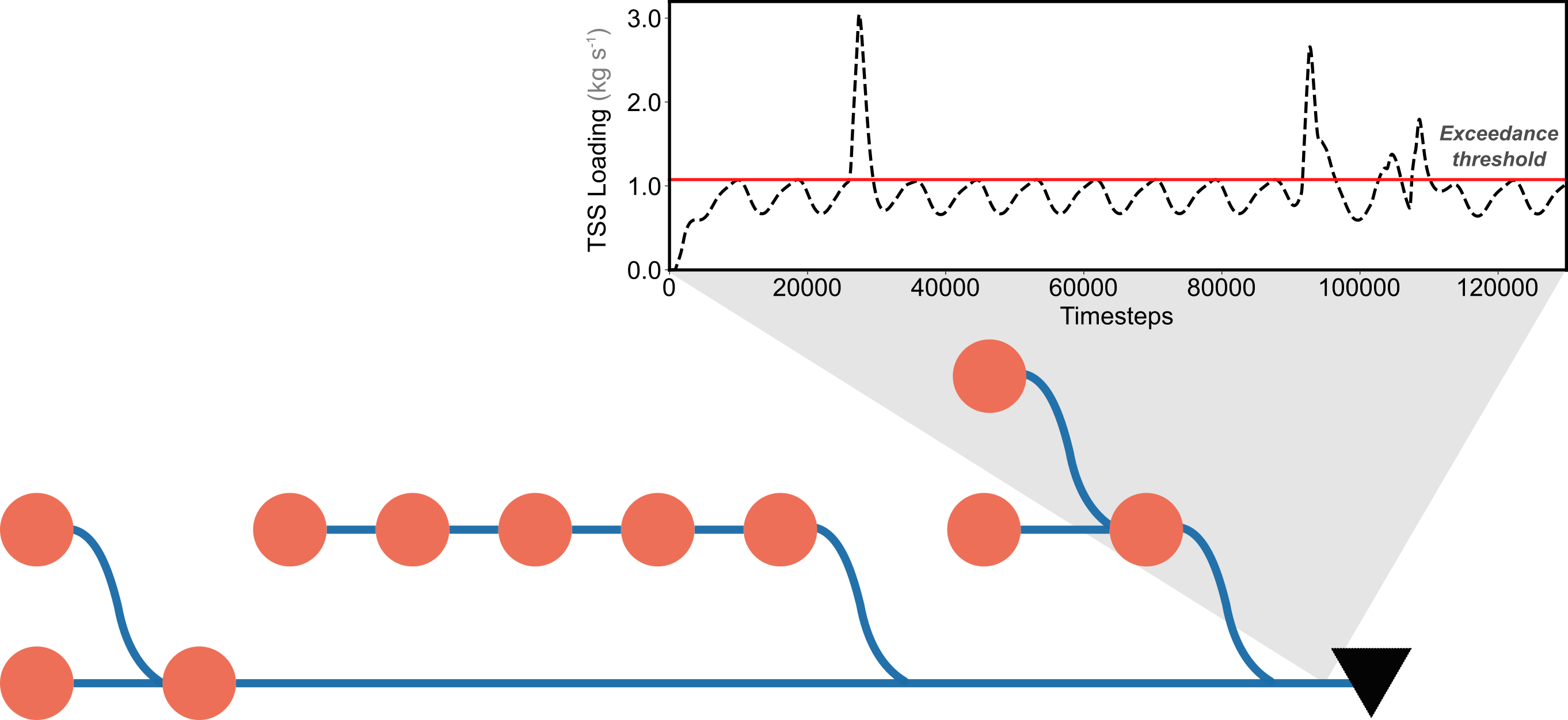}
    \caption{Topology for the separated combined sewer network of Scenario \scenario{epsilon}.
    Orange circles and blue lines in the figure represent respectively the controllable storage elements and channels draining into the outlet (black triangle).
    During a storm event, each node in the network receives runoff generated by connected subcatchments in addition to discharge from upstream nodes.
    The system's response to the given storm event for the uncontrolled case (i.e. all 11 of its in-line storage dams are set to 100\% open) is depicted via plotting the outlet's TSS load against the target exceedance threshold for each timestep of the simulation.
    For this scenario, the controller is penalized for any TSS load above given exceedance thresholds for the downstream outlet.}
    \label{fig:network_epsilon}
\end{figure}
\begin{table}[ht]
\small
\caption{Scenario \scenario{epsilon} Configuration}
\begin{tabular}{l p{5.5in}}
\toprule
\textbf{Network Topology} & 
Scenario \scenario{epsilon} is a combined sewer system in an urbanized watershed of $\approx 67\ km^2$. 
Scenario \scenario{epsilon}'s network is comprised of 11 in-line storage dams, connected in both series and parallel, that discharge to the downstream water resource recovery facility (WRRF). The stormwater network topology for Scenario \scenario{epsilon} can be seen in Figure \ref{fig:network_epsilon}. \\\midrule
\textbf{Event Driver} & 
Scenario \scenario{epsilon} is driven by a 4- day period of measured precipitation data collected from 7 rain gauges located across the network. 
Diurnal wastewater inflows are also included at the receiving nodes for each subcatchment. The magnitude of these inflows is proportional to the area of each subcatchment. \\\midrule
\textbf{Controllable Assets} & 
Each of Scenario \scenario{epsilon}'s in-line storage dams are control assets modeled as weirs with adjustable settings within a range of (0\% to 100\%) during the simulation. 
The percentage of each weir corresponds to the percent opening of the corresponding storage dam, with all valves initially set to 100\% open. 
In the computational implementation of Scenario \scenario{epsilon}, each of the controllable valves are identified with the pattern \texttt{ISD<storage asset ID>} (e.g. ISD001, ISD002, \ldots ISD011).\\\midrule
\textbf{Observable States} & 
The observable states for Scenario \scenario{epsilon} are the outflow, water levels, and TSS concentrations for all in-line storage dams in the network, which can be queried throughout the simulation to inform any control decisions.
The outflow and TSS concentration at the outlet of the network are observable as well. \\\midrule
\textbf{Control Objectives} & 
The primary objective for Scenario \scenario{epsilon} is to maintain the TSS load at the outlet of the sewer network below a given threshold ($1.05\ kg/s$), which is the maximum dry-weather TSS load observed, while also avoiding flooding. The ability to achieve Scenario \scenario{epsilon}'s objective is quantified via its performance, $P$, in the equation below. As seen from the equation, flooding constitutes a much greater penalty than threshold exceedance. Thus, it is presumed a control algorithm will permit threshold exceedance when necessary to avoid flooding.
\begin{math}
\begin{aligned}[t]
P(t) &= \begin{cases}
0.0, & TSS(t) \leq 1.05 \quad and \quad  F(t) = 0 \\
TSS(t) - 1.05, & TSS(t) > 1.05 \quad and \quad  F(t) = 0 \\
10^6, & F(t) > 0
\end{cases}\\
P &= \sum^T_{t=0} P(t)
\end{aligned}
\end{math}
where $TSS(t)$ and  $F(t)$ represent the TSS load at the network outlet and overall network flooding, respectively, for time $t$.
\\ \midrule
\textbf{Additional Notes} & Interactive example: \demoepsilon\\
\bottomrule
\end{tabular}
\end{table}
\clearpage
\subsection{Scenario \scenario{theta}}
\label{si:theta}
\begin{figure}[ht!]
    \centering
    \includegraphics[width=0.50\textwidth]{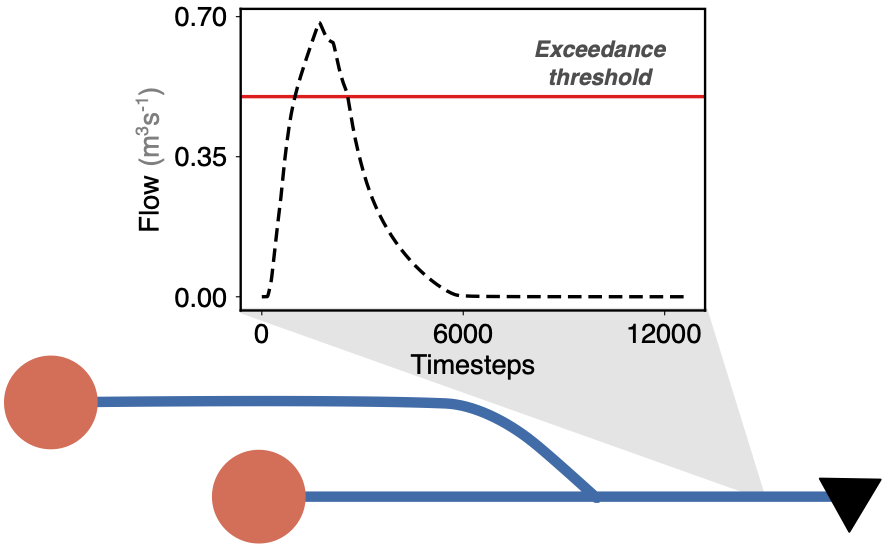}
    \caption{Topology for the separated stormwater network of Scenario \scenario{theta}.
	Orange circles and blue lines in the figure represent respectively the controllable storage elements and channels draining into the outlet (black triangle). 
	During a storm event, each node in the network receives runoff generated by connected sub-catchments. 
	The system's response to the given storm event for the uncontrolled case (i.e. all 2 of its valves set to 100\% open) is depicted via plotting the outlet's flows against the target exceedance threshold for each timestep of the simulation. 
	For this scenario, the controller is penalized for any flow rates above given exceedance threshold at the network outlet and causing flooding in the network. }\label{fig:network_theta}
\end{figure}
\begin{table}[ht]
\small
\caption{Scenario \scenario{theta} Configuration}
\begin{tabular}{l p{5.5in}}
\toprule
\textbf{Network Topology} & 
Scenario \scenario{theta} is a simple network is designed for unit testing of control algorithms. It consists of two storage basins with a controllable outlet that discharges to an outfall. The stormwater network topology for Scenario \scenario{theta} can be seen in Figure. \ref{fig:network_theta}. \\\midrule
\textbf{Event Driver} & 
This scenario is driven by synthetic storm event of 3 inches per hour peak intensity and distributed uniformly throughout the entire subcatchment. The uncontrolled response of the system at the outlet is illustrated in the Figure. \ref{fig:network_theta}.\\\midrule
\textbf{Controllable Assets} & 
The two tanks in this scenario are equipped with 1 $m^2$ square orifice.
These orifice's opening percentage (e.g., 10\%, 73\%) can be adjusted throughout the duration of the simulation.
These orifices are initially set to 100\% open. In the computational implementation of Scenario \scenario{theta}, these control assets are identified as \texttt{1} and \texttt{2}. \\\midrule
\textbf{Observable States} & 
The observable states for Scenario \scenario{theta} are the outflow $[m^3/s]$, water level $[m]$, and valve setting $[\%]$ for the tank and orifice elements in network, which can be queried throughout the simulation to inform any control decisions. \\\midrule
\textbf{Control Objectives} & 
Control objective Scenario \scenario{theta} is to maintain the outflows at the outlet below the desired threshold ($0.5 m^3/s$) while also avoiding flooding.
The ability to achieve Scenario \scenario{theta}'s objective is qualified via its performance metric, $P$, summarized below. As seen in the equation, flooding constitutes a much greater penalty than threshold exceedance.
\begin{math}
\begin{aligned}[t]
P_i(t) &= \begin{cases}
0.0, & Q_i(t) \leq 0.50 \quad and \quad  F_i(t) = 0 \\
Q_i(t) - 0.50, & Q_i(t) > 0.50 \quad and \quad  F_i(t) = 0 \\
10^6, & F_i(t) > 0
\end{cases}\\
P &= \sum^{2}_{i=1} \sum^T_{t=0} P_i(t)
\end{aligned}
\end{math}
Where $Q_i(t)$ and $F_i(t)$ represents the outflow and flooding, respectively, at the $i^{th}$ node of the network for time $t$.
\\ \midrule
\textbf{Additional Notes} & Interactive example: \demotheta\\
\bottomrule
\end{tabular}
\end{table}
\clearpage
\subsection{Scenario \scenario{zeta}}
\label{si:zeta}
\begin{figure}[ht!]
    \centering
    \includegraphics[width=0.8\textwidth]{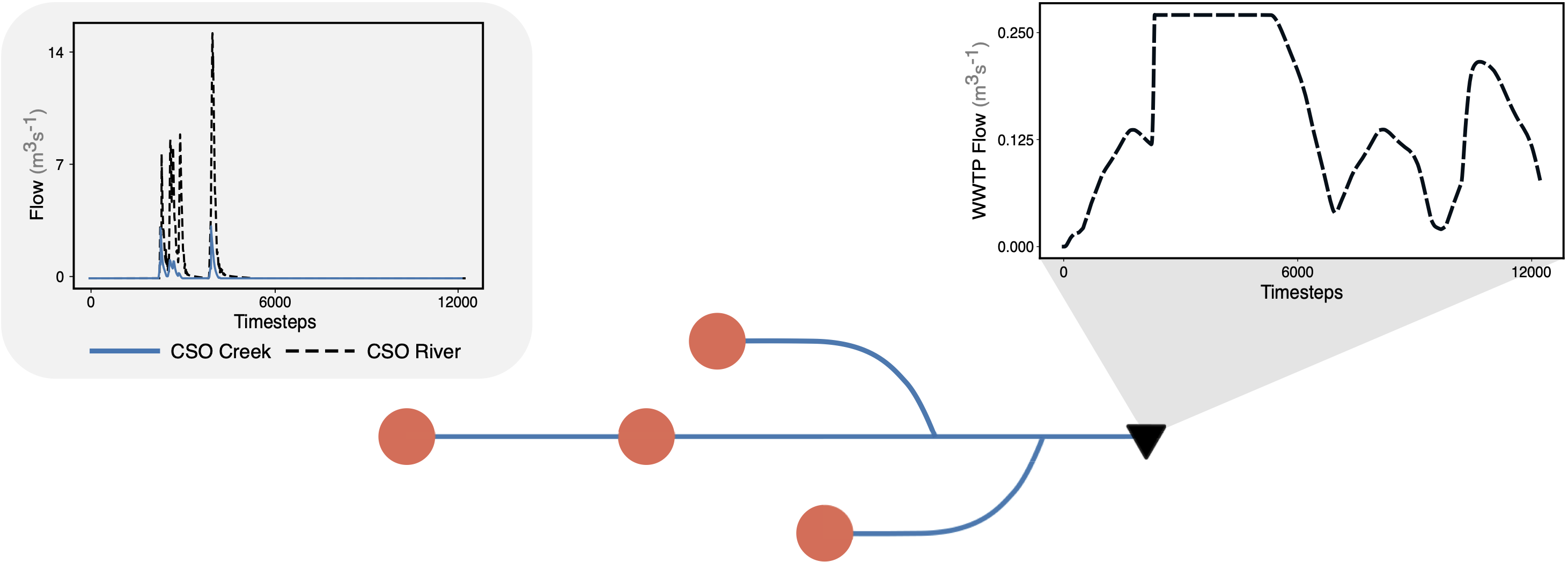}
    \caption{Topology for the combined network of Scenario \scenario{zeta}.
	Orange circles and blue lines in the figure represent, respectively, the controllable storage elements and channels draining the outflow (black triangle) of the subnetwork. 
	During a storm event, each storage element in the network receives runoff generated by connected subcatchments, in addition to discharge from upstream storage elements. 
	The system's response to the given storm event for the uncontrolled case (i.e. all 4 of its valves set to 100\% open) is depicted  for each timestep of the simulation by plotting the combined sewer overflow into the receiving creek and river, and the network outflow to the downstream wastewater treatment plant.
	For this scenario, the controller is penalized for any combined sewer overflow events, and rewarded for the outflow volume to the wastewater treatment plant.}\label{fig:network_zeta}
\end{figure}
\begin{table}[ht]
\small
\caption{Scenario \scenario{zeta} Configuration}
\begin{tabular}{l p{5.5in}}
\toprule
\textbf{Network Topology} & 
Scenario \scenario{zeta} is a combined system with ten subcatchments covering an area of $\approx 1.8 \unit{km^2}$, with six storage tanks (a total storage volume of $5900 \unit{m^3}$) connected both in series and parallel, and discharging to the downstream wastewater treatment plant. Within the network, there are ten combined sewer overflow (CSO) structures (one at each storage basin, plus an additional four locations), which drain into two receiving bodies: a  river and a creek. The combined sewer overflows of Tank 1, Tank 2, Tank 3, Tank 4, Tank 5, CSO8, and CSO10 overflow into the larger river, while Tank 6, CSO7, and CSO9 overflow into the smaller, more sensitive creek which flows through a local park. Of the six storage basins, four are equipped with controllable outlet valves to utilize system storage and minimize the volume and number of combined sewer overflow events (Tank 2, Tank 3, Tank 4, and Tank 6).
\vspace{0.5em}
\newline
Scenario \scenario{zeta} is based on the SWMM implementation of the Astlingen Real-Time Control Benchmarking Network \citep{Sun2020}. The topology of Scenario \scenario{zeta}, specifically the controllable tanks, can be seen in Figure \ref{fig:network_zeta}. 
\\\midrule
\textbf{Event Driver} & 
Rainfall in Scenario \scenario{zeta} is spatially heterogeneous, with four rainfall gauges connected to the ten subcatchments. 
Precipitation for the original Astlingen network from \citet{Sun2020} was for over a  10-year period. However, Scenario \scenario{zeta} uses one precipitation event from this 10-year data as its event driver.
Specifically, Scenario \scenario{zeta} is driven by an 3-day precipitation event, with the majority of precipitation occurring over a 12-hr period, averaging $53.06 \unit{mm}$ of rainfall across four rain gages, characterizing this event as a 17-year storm. (Note, this event was characterized using annual recurrence interval and rainfall duration data for a similar geographic location presented in \citet{Ramke2018}).   
\\\midrule
\textbf{Controllable Assets} & 
Four of Scenario \scenario{zeta}'s storage tanks are equipped with valves at their outlets and adjustable to a setting within a range of (0\% to 100\%) during the simulation. The controllable valves are implemented as orifices and are identified as \texttt{<}\texttt{V2}, \texttt{V3}, \texttt{V4}, \texttt{V6}\texttt{>}, corresponding to tanks \texttt{<}\texttt{T2}, \texttt{T3}, \texttt{T4}, \texttt{T6}\texttt{>}, respectively. 
\\\midrule
\textbf{Observable States} & 
The observable states for Scenario \scenario{zeta} are the water levels at the six storage tanks: \texttt{<}\texttt{T1}, \texttt{T2}, \texttt{T3}, \texttt{T4}, \texttt{T5}, \texttt{T6}\texttt{>}, and can be queried throughout the simulation to inform any control decisions. \\\midrule
\textbf{Control Objectives} & 
The primary objective for Scenario \scenario{zeta} is to minimize combined sewer overflow into the receiving river and creek, and maximize the downstream flow to the wastewater treatment plant. As the creek is more sensitive to overflows than the river, it is desirable to minimize overflow to the creek over that of the river. Additionally, there is a desire to minimize the roughness of control actions. The ability to achieve Scenario \scenario{zeta}'s objective is quantified via its performance, $P$, in the following equation: 
\begin{center}
\begin{math}
\begin{aligned}[t]
P = \sum^T_{t=0} \left[\left(\sum^{7}_{i=1} V_{\mathrm{CSO}_i}(t)\right) +
    2\left(\sum^{3}_{j=1} V_{\mathrm{CSO}_j}(t)\right) +
    0.01\left(\sum^{6}_{k=1} V_{\mathrm{throttle}_k}(t)\right) -
    0.1V_{\mathrm{WWTP}}(t)\right]
\end{aligned}
\end{math}
\end{center}
where $V_{\mathrm{CSO}_i}(t)$ is the CSO volume at the $i$ locations of \texttt{<}\texttt{T1}, \texttt{T2}, \texttt{T3}, \texttt{T4}, \texttt{T5}, \texttt{CSO8}, \texttt{CSO10}\texttt{>} that overflow into the river; $V_{\mathrm{CSO}_j}(t)$ is the CSO volume at the $j$ locations of \texttt{<}\texttt{T6}, \texttt{CSO7}, \texttt{CSO9}\texttt{>} that overflow into the creek; $V_{\mathrm{throttle}_k}(t)$ is the change volume change in flow at each of the controlled and uncontrolled $k$ tank outlets \texttt{<}\texttt{V1}, \texttt{V2}, \texttt{V3}, \texttt{V4}, \texttt{V5}, \texttt{V6}\texttt{>}; and $V_{\mathrm{WWTP}}(t)$ is the volume of wastewater flowing to the downstream wastewater treatment plant, all over timestep $t$.  
\\ \midrule
\textbf{Additional Notes} & 
As stated above, Scenario \scenario{zeta} is based on a SWMM implementation of the Astlingen Real-Time Control Benchmarking Network, which was originally developed by the Integral Real-Time Control working group of the German Water Association for the purpose of comparing control strategies applied to the network \citep{Sun2020}. Three different control strategies were presented and tested on the original SWMM implementation of the Astlingen network. We implement two of these control strategies---a rule-based and equal-filling degree approach---to Scenario \scenario{zeta} in the following interactive example: \demozeta.\\
\bottomrule
\end{tabular}
\end{table}
%
%
%
%
\end{document}